\documentclass[prd,reprint,nofootinbib,superscriptaddress,preprintnumbers,showpacs]{revtex4-1}
\usepackage{amsmath}	
\usepackage{amssymb}	
\usepackage{amstext}	
\usepackage{latexsym}	
\usepackage{bm}			
\usepackage{graphicx,color}	
\usepackage{slashed}	
\usepackage{makecell}	
\usepackage{multirow}	
\usepackage{caption}	
\usepackage{hyperref}	
\usepackage{amsthm,amsmath,amssymb}
\usepackage{mathrsfs}

\allowdisplaybreaks[4]
\pdfoutput=1

\begin{document}
	\title{The Theoretical Study of $\Sigma^{+} p \to\Lambda a_0^{+} p$ Reaction}
	\date{\today}
	\author{Yan Dai}
	\affiliation{School of Physics and Optoelectronics Engineering, Anhui University, Hefei 230601, People's Republic of China}
	\author{Aojia Xu}
	\affiliation{School of Physics, Dalian University of Technology, Dalian 116024, People's Republic of China}
	\author{Gang Li}
	\affiliation{School of Physics and Optoelectronics Engineering, Anhui University, Hefei 230601, People's Republic of China}
	\author{Mao Song}
	\affiliation{School of Physics and Optoelectronics Engineering, Anhui University, Hefei 230601, People's Republic of China}
	\author{Xuan Luo}
	\email{xuanluo@ahu.edu.cn}
	\affiliation{School of Physics and Optoelectronics Engineering, Anhui University, Hefei 230601, People's Republic of China}
	\begin{abstract}
		We conducted a theoretical study on the process $\Sigma^{+} p \to\Lambda a_0^{+} p$ based on an effective Lagrangian approach. This model encompasses the excitation of intermediate states leading to the production of $\Delta(1920)$ through $\pi^{+}$ and $K^{+}$ meson exchanges between the initial $\Sigma^{+}$ baryon and the initial proton $p$, as well as the generation of $\Delta(1940)$ via $\pi^{+}$ and $\rho^{+}$ meson exchanges. We provide predictions for the total and differential cross sections and discuss the potential impacts of cutoff parameters, off-shell effects, and branching ratios on the $\Delta^{\ast} \to a_0 p$ decay. Given the dominance of $\Delta(1940)$ in the region close to the reaction threshold, this reaction is considered an ideal platform for deeply exploring the unique properties of the $\Delta(1940)$ resonance. Through this approach, we can gain a more precise understanding of the intrinsic characteristics of this particle.
	\end{abstract}
	\maketitle
	\section{introduction}
   The study of meson-baryon interactions at low energies is a significant topic at the intersection of particle physics and nuclear physics, offering immense value for a deeper understanding of the fundamental constituents of matter and the laws governing their interactions. Within this expansive research field, the exploration of hyperon resonances holds a particularly prominent position. Hyperons, a class of baryons that contain strange quarks (s-quarks), enhance our comprehension of the material world through their unique properties and provide a valuable experimental and theoretical platform for testing and developing Quantum Chromodynamics (QCD), the fundamental theory that describes strong interactions.
	
   Within the field of hyperon resonance research, particularly focusing on the $\Delta(1920)$ and $\Delta(1940)$ hyperon resonances ~\cite{Ronchen:2022hqk,Hunt:2018wqz,CBELSATAPS:2015kka,Ronchen:2015vfa,CBELSATAPS:2014wvh,Svarc:2014zja,Shrestha:2012ep,Anisovich:2011fc,CB-ELSA:2008zkd,Penner:2002ma,Penner:2002md,Vrana:1999nt,Hohler:1993lbk,Cutkosky:1980rh,Hohler:1979yr}, experimental data on hyperon resonances are relatively scarce compared to the extensive research available in the field of nucleon resonances, such as the well-studied $\Delta(1232)$. However, valuable information about $\Delta^{\ast}$ hyperon resonances can still be obtained through photon-induced reactions ~\cite{CLAS:2018drk,CBELSATAPS:2021osa,CLAS:2023akb,Kohri:2006yx,CBELSATAPS:2015kka,CBELSATAPS:2014wvh} and N$\pi$ scattering experiments ~\cite{CBELSATAPS:2015kka,CBELSATAPS:2014wvh,Ronchen:2015vfa,Anisovich:2011fc,Ronchen:2022hqk}. Notably, in the $\gamma p\to\pi^{0}\eta$ reaction , we observed for the first time the existence of parity doublets with total angular momentum $J = \frac{3}{2}$ for the $\Delta(1920)$ and $\Delta(1940)$ resonances , these resonances can decay into an $a_0(980)$ pair~\cite{CB-ELSA:2007xbv,Wang:2023lnb}.
   Despite this, current experiments and research on $\Delta^{\ast}$ resonances coupled with the $a_0(980)p$ channel remain relatively scarce. Overall, there is an urgent need to strengthen experimental research on $\Delta^{\ast}$ hyperon resonances to address this knowledge gap and advance the development of related theories. This situation primarily arises from the challenging conditions required for the production of hyperon resonances, which not only necessitate specific reaction environments but also impose relatively high technical thresholds for experimentation. Furthermore, hyperons are unstable and tend to decay rapidly after their production, which undoubtedly increases the difficulty and complexity of experimental observations. These studies not only elucidate the fundamental properties of $\Delta$ hyperon resonances (including $\Delta^{\ast}$), such as mass, width, spin, and decay patterns, but also provide valuable insights into the interaction mechanisms between hyperons and mesons.
   In our current research, we have specifically introduced the reaction $\Sigma^{+} p \to \Lambda a_0^{+} p$, which, due to its unique properties, serves as an ideal platform for exploring $\Delta^{\ast}$ resonances with isospin J=$\frac{3}{2}$. This reaction effectively eliminates interference from $N^{\ast}$ resonances with isospin $\frac{1}{2}$ through the principle of charge conservation, thereby ensuring the integrity of our research. Consequently, we focus on the direct coupling effect between intermediate resonant states and the final state $a_0 p$ pair, avoiding interference from unrelated channels. Since we have not observed significant decay signals of $\Delta^{\ast}$ resonances leading to other pathways involving $\Lambda$ particles within the energy range of interest, this characteristic allows us to investigate more clearly the interaction between $\Delta^{\ast}$ resonances and $a_0 p$ pairs.
   Furthermore, the threshold energy of the $a_0p$ channel, approximately 1.918~GeV, closely aligns with the mass ranges of the $\Delta(1920)$ and $\Delta(1940)$ resonances. This correlation significantly enhances the potential of this reaction to reveal the properties of these resonant states. Consequently, this reaction not only deepens our understanding of fundamental reaction mechanisms but also offers a valuable experimental opportunity to explore the coupling mechanisms between $\Delta^{\ast}$ resonances and exchanged particles.
   In this study, we specifically propose the use of the $\Sigma^{+} p \to \Lambda a_0^{+} p$ reaction to investigate the characteristics of the $\Delta(1940)$ resonance. We employ an effective Lagrangian approach to analyze this reaction, with a particular focus on the production of the $\Delta(1940)$ hyperon resonance. Effective Lagrange methods are widely used in calculating reaction cross sections to explore the reaction mechanisms between initial particles and final particles in particle collisions.~\cite{Xu:2024tiu,Zhang:2018kdz,Xie:2014kja,Cheng:2016hxi,Liu:2019ojr,Wang:2014ofa,Oh:2007jd}. Given the large decay branching ratio of the $\Delta(1940)$ to the $N\rho$ channel, we speculate that there is a significant coupling between the $\Delta(1940)$ and the $N\rho$ channel. Therefore, we expect $\rho$ meson exchange will play a crucial role in exciting the $\Delta(1940)$ resonance in this reaction. In our model, the $\Delta(1920)$ and $\Delta(1940)$ resonances are excited through the exchanges of $\pi^{+}$ and $K^{+}$, as well as $\rho^{+}$ and $\pi^{+}$ mesons, respectively, between the initial $\Sigma^{+}$ particle and the proton $p$. This research not only enhances our understanding of the properties of $\Delta^{\ast}$ resonances but also provides new directions and insights for future experimental research and theoretical exploration.
   Our work presents predictions for total cross-sections, angular distributions, and invariant mass distributions, which will facilitate future comparisons with experimental data. We also discuss the potential impacts of cutoff parameters, off-shell effects, and branching ratios on the decay of $\Delta^{\ast}\to a_0p$, specifically regarding both total and differential cross-sections. The organization of our work is as follows: In Sec.$\mathrm{II}$, we introduce the formalism and necessary components of each amplitude in the model and derive the specific forms of the amplitudes. The numerical results, including the summed and differential potential cross-sections for the reaction $\Sigma^{+} p\to\Lambda a_0^{+}p$, are presented in Sec.$\mathrm{III}$. Finally, a brief conclusion is drawn in Sec.$\mathrm{IV}$.
	
	\section{Formalism}
\begin{figure}[htpb]
		\centering
		\includegraphics[width=0.4\textwidth]{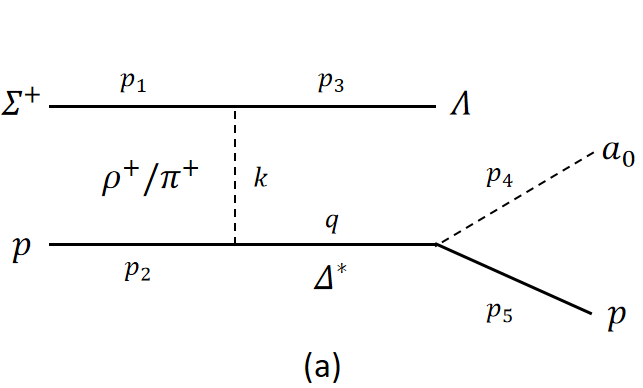}\hypertarget{1a}{}\\
		\includegraphics[width=0.4\textwidth]{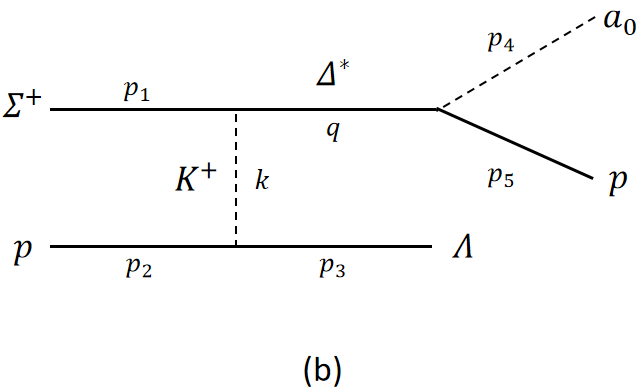}
		\captionsetup{justification=raggedright}
		\caption{(a) $t$-(b) $u$-channel exchanges Feynman diagrams for $\Sigma^{+} p\to\Lambda a_0^{+}p$.}
		\label{1}
	\end{figure}

	In the effective Lagrangian approach, the production mechanisms of $\Delta(1920)$ and $\Delta(1940)$ resonances in the reaction $\Sigma^{+} p\to\Lambda a_0^{+}p$ consist of standard $t$- and $u$-channel components, with the $\rho^{+}$
	and $\pi^{+}$ meson serving as the $t$-channel exchange, as depicted in Fig.~\ref{1}. Given the strong coupling between $\Delta(1940)$ and the $N\rho$ channel, we anticipate that it may play a significant role in the reaction. Due to charge considerations, $\pi$ and $\rho$ exchanges are exclusive to charged $\pi^{+}$ and $\rho^{+}$, respectively. Similarly, as $\Delta(1920)$ exhibits a strong coupling with the $N\pi$ channel, $\pi$ and $K$ exchanges are limited to charged $\pi^{+}$ and $K^{+}$.
	
	The production amplitude is calculated from the following effective Lagrangians~\cite{Penner:2002ma,Nakayama:2006ty,Huang:2012xj,Cheng:2016hxi},
	\begin{align}
		&\mathcal{L}_{\pi\Lambda\Sigma}=-\frac{g_{\pi\Lambda\Sigma}}{m_{\Lambda}+m_{\Sigma}}\bar{\Lambda}\gamma_5 \slashed{\partial}\pi\cdot\Sigma ,\\
		&\mathcal{L}_{K\Lambda N}=-\frac{g_{K\Lambda N}}{m_{\Lambda}+m_p}\bar{\Lambda}\gamma_5\slashed{\partial}\bar{K}N+h.c. ,\\
		&\mathcal{L}_{\rho\Lambda\Sigma}=-g_{\rho\Lambda\Sigma}\bar{\Lambda}(\gamma^{\mu}-\frac{k_{\rho\Lambda\Sigma}}{m_{\Lambda}+m_{\Sigma}}(k^{\mu}_{\rho}-\slashed{k}_{\rho}\gamma^{\mu}))\rho_{\mu}\cdot\Sigma .
	\end{align}
	for $\Delta(1920)$,
	\begin{align}
		&\mathcal{L}_{\Delta^{\ast} Na_0}=-\frac{ig_{\Delta^{\ast} Na_0}}{m_{a_0}}\bar{N}\partial^{\mu}\tau\cdot a_0\gamma_5\Theta_{\mu\nu}(Z1)\Delta^{\ast\nu}+h.c. ,\\
		&\mathcal{L}_{\Delta^{\ast} N\pi}=\frac{{g_{\Delta^{\ast} N\pi}}}{m_{\pi}}\bar{\Delta}^{\ast\mu}\Theta_{\mu\nu}(Z1)\partial^{\nu} \tau\cdot\pi N+h.c. ,\\
		&\mathcal{L}_{\Delta^{\ast} K\Sigma}=\frac{f_{\Delta K\Sigma}}{m_K}\bar{\Delta}^{\ast\mu}\Theta_{\mu\nu}(Z1)(\partial^{\nu} K)\Sigma+h.c.,
	\end{align}
	for$\Delta(1940)$,
	\begin{align}
		&\mathcal{L}_{\Delta^{\ast} Na_0}=-\frac{g_{\Delta^{\ast} Na_0}}{m_{a_0}}\bar{N}\partial^{\mu}\tau\cdot a_0\Theta_{\mu\nu}(Z2)\Delta^{\ast\nu}+h.c., \\
		&\mathcal{L}_{\Delta^{\ast} N\rho}=-i\frac{g_{\Delta^{\ast} N\rho}}{2m_p}\bar{\Delta}^{\ast}_{\mu}\gamma_{\nu}\rho^{\mu\nu}N+h.c. ,\\
		&\mathcal{L}_{\Delta^{\ast} N\pi}=-i\frac{{g_{\Delta^{\ast} N\pi}}}{m_{\pi}}\bar{\Delta}^{\ast\mu}\gamma_5\Theta_{\mu\nu}(Z2)\partial^{\nu} \tau\cdot\pi N+h.c.,
	\end{align}
	with
	\begin{equation}
		\Theta_{\mu\nu}(Z)=g_{\mu\nu}-(Z+\frac{1}{2})\gamma_\mu\gamma_\nu.
	\end{equation}
	The introduction of off-shell parameters Z~\cite{Benmerrouche:1989uc,Mizutani:1997sd,Penner:2002ma} in the effective Lagrangian framework plays a pivotal role in describing the off-shell effects of high-spin particles, enhancing the accuracy of theoretical models, and bridging the gap between theory and experiment. In this context, the off-shell parameters $\Theta_{\mu\nu}(Z1)$ and $\Theta_{\mu\nu}(Z2)$ in the Lagrangian correspond to $\Delta(1920)$ and $\Delta(1940)$ respectively. Here, $\kappa_{\rho\Lambda\Sigma} = -30.50$ ~\cite{Ronchen:2012eg} represents the anomalous magnetic moment. The coupling constants can be predicted based on SU(3) symmetry, yielding $g_{\pi\Lambda\Sigma} = 0.69$~GeV$^{-1}$, $g_{K\Lambda N} = -1.03$~GeV$^{-1}$, and $g_{\rho\Lambda\Sigma} = -0.56$~GeV$^{-1}$ ~\cite{Ronchen:2012eg}.
	
	Other coupling constants are determined from partial decay widths, as shown in Table 1. We extracted the coupling constants related to vertices involving $\Delta^{\ast}$ from the decay widths. However, this approach does not allow us to determine the relative phase between $\Delta(1920)$ and $\Delta(1940)$. Therefore, we considered multiplying by a factor of -1 to detect the presence of interference effects. When studying interference effects, no significant variations were observed for either $\Delta(1920)$ or $\Delta(1940)$. Notably, we used the average branching ratios listed in the Particle Data Group ~\cite{ParticleDataGroup:2022pth}. Given that $g^{(2)}_{\Delta(1940)N\rho}$ and $g^{(3)}_{\Delta(1940)N\rho}$ in the $\Delta(1940)N\rho$ interaction have never been rigorously calculated, we only computed $g^{(1)}_{\Delta(1940)N\rho}$ through the $\Delta(1940)\to N\rho$ decay, neglecting $g^{(2)}_{\Delta(1940)N\rho}$ and $g^{(3)}_{\Delta(1940)N\rho}$ ~\cite{Wang:2018vlv}. Since the mass of $\Delta(1920)$ is very close to the $a_0p$ threshold, their finite widths must be taken into account. We use the following formula to include the finite width effects~\cite{Wang:2023lnb,Roca:2005nm}:
	\begin{equation}
		\begin{split}
			\Gamma_{\Delta^*\to Na_0}=&-\dfrac{1}{\pi}\int_{(m_{\Delta^*}-2\Gamma_{\Delta^*})^2}^{(m_{\Delta^*}+2\Gamma_{\Delta^*})^2}ds\Gamma_{\Delta^*\to Na_0}(\sqrt{s})\\
			&\Theta(\sqrt{s}-m_N-m_{a_0})\mathrm{Im}\left\{\dfrac{1}{s-m_{\Delta^*}^2+im_{\Delta^*}\Gamma_{\Delta^*}}\right\}.
		\end{split}
	\end{equation}
	\begin{table}[htbp]
		\renewcommand{\arraystretch}{1.8}
		\tabcolsep=1.6mm
		\captionsetup{justification=raggedright}
		\caption*{TABLE~I. Coupling constants used in this work.}
		\hypertarget{tab1}{}
		\begin{tabular}[b]{ccccc}
			State & \makecell{Width\\(MeV)} & \makecell{Decay\\channel} & \makecell{Branching ratio\\adopted} & $g^2/4\pi$\\
			$\Delta(1920)$ & 300 & $N\pi$ & 0.125  & $5.26\times 10^{-3}$\\
			&  & $\Sigma K$ & 0.04  & $8.92\times 10^{-2}$\\
			&  & $Na_0(980)$ & 0.04  & $4.12$\\
			$\Delta(1940)$ & 400 & $N\rho$ & 0.8  & $5.39$\\
			&  & $N\pi$ & 0.105  & $3.85\times 10^{-2}$\\
			&  & $Na_0(980)$ & 0.02  & $1.06$
		\end{tabular}
	\end{table}
	
	Since hadrons are not point-like particles, here we need to consider the shape factor of each vertex, which can be used to parametrically describe the internal structure of hadrons. Next, we introduce the form factors for the various meson exchanges. For $K$ mesons and $\pi$ mesons, n=1 here; for $\rho$ mesons, n=2 here,
	\begin{equation}
		f_M(k_M^2)=\left(\dfrac{\Lambda_M^2-m_M^2}{\Lambda_M^2-k_M^2}\right)^n.
	\end{equation}
	In the equation, $k_M$ and $m_M$ denote the four-momentum and mass of the exchanged meson, respectively. Here, we use $\Lambda_{\rho} = 1.2 \, \text{GeV}$~\cite{Fan:2019lwc} to represent the exchange of the $\rho$ meson. For the intermediate baryons, we introduce a form factor,
	\begin{equation}
		f_B(q_B^2)=\left(\dfrac{\Lambda_B^4}{\Lambda_B^4+(q_B^2-m_B^2)^2}\right)^2.
	\end{equation}
	The four-momentum and mass of the intermediate baryons are denoted by $q_B$ and $m_B$, respectively. For the cutoff parameters of the intermediate baryons, we take $\Lambda_{\Delta(1920)} = \Lambda_{\Delta(1940)} = 1.5 \, \text{GeV}$.
	
	The propagators for the exchanged particles are expressed as
	\begin{equation}
		G_\rho^{\mu\nu}(k)=-i\left(\dfrac{g^{\mu\nu}-\frac{k^\mu k^\nu}{m_\rho^2}}{k^2-m_\rho^2}\right),
	\end{equation}
	for $\rho$ meson,
	\begin{equation}
		G^0(q)=\frac{i}{(q^2-m_M^2)},
	\end{equation}
	for $K$,$\pi$ meson,
	\begin{equation}
		G_p(k)=\dfrac{i(\slashed{k}+m_p)}{k^2-m_p^2},
	\end{equation}
	for $p$ baryon, and
	\begin{equation}
		\begin{split}
			G^\frac{3}{2}(q)=&\dfrac{i(\slashed{q}+M_R)}{q^2-M_R^2+iM_R\Gamma_R}\left[g_{\mu\nu}-\dfrac{1}{3}\gamma_\mu\gamma_\nu\right.\\
			&\left.-\dfrac{1}{3M_R}(\gamma_\mu q_\nu-\gamma_\nu q_\mu)-\dfrac{2}{3M_R^2}q_\mu q_\nu\right].
		\end{split}
	\end{equation}
	For baryons with a spin-3/2, where $k$ and $q$ represent the four-momenta of the exchanged particles, and $M_R$ and $\Gamma_R$ denote the mass and width of the intermediate resonance, respectively.
	Under the aforementioned conditions, the total scattering amplitude for the reaction $\Sigma^+ p \rightarrow\Lambda a_0^{+} p$ can be expressed as:
	\begin{equation}
		\begin{split}
			\mathcal{M}^{\pi}_a=
			&\frac{16ig_{\Delta^{\ast} Na_0}g_{\Delta^{\ast}N\pi}g_{\pi\Lambda\Sigma}}{(m_{\Lambda}+m_{\Sigma})m_{\pi}m_{a_0}}f_{\pi}(k_1)f^2_{\Delta^{\ast}}(q)\bar{u}(p_5,s_5)(p_{a_0}^{\mu})\\
			&\gamma_5\Theta_{\mu\nu}(Z1)G^{\nu\rho}_{\Delta^{\ast}}\Theta_{\rho\sigma}(Z1)(p_{\pi}^{\sigma})u(p_2,s_2)G_{\pi}(p_{\pi})\\
			&\bar{u}(p_3,s_3)\gamma_5(\slashed{p}_{\pi})u(p_1,s_1) ,\\
			\mathcal{M}^K_b=
			&-\frac{8\sqrt{2}g_{\Delta^{\ast} Na_0}f_{\Delta^{\ast}\Sigma K}g_{K\Lambda N}}{m_{a_0}m_K(m_{\Lambda}+m_p)}f_{K}(k_2)f^2_{\Delta^{\ast}}(q)\bar{u}(p_5,s_5)(p^{\mu}_{a_0})\\
			&\gamma_5\Theta_{\mu\nu}(Z1)G^{\nu\rho}_{\Delta^{\ast}}\Theta_{\rho\sigma}(Z1)p^{\sigma}_Ku(p_1,s_1)G_{K}(p_K)\bar{u}(p_3,s_3)\\
			&\gamma_5\slashed{k}_1u(p_2,s_2) ,\\
			\mathcal{M}^{\rho}_c=
			&\frac{8\sqrt{2}g_{\Delta^{\ast} Na_0}g_{\Delta N\rho}g_{\rho\Lambda\Sigma}}{2m_{a_0}m_p}f_{\rho}(k_1)f^2_{\Delta^{\ast}}(q)\bar{u}(p_5,s_5)(p^{\mu}_{a_0})\\
			&\Theta_{\mu\nu}(Z2)G^{\nu\rho}_{\Delta^{\ast}}\gamma_{\sigma}(k_{1\rho}G_{\rho}^{\sigma\eta}(k_1)-k_{1\sigma}G_{\rho}^{\rho\eta}(k_1))u(p_2,s_2)\\
			&\bar{u}(p_3,s_3)(\gamma_{\eta}+\frac{k_{\rho\Lambda\Sigma}}{m_{\Lambda}+m_{\Sigma}}(p_{\rho\eta}-\slashed{p}_{\rho}\gamma_{\eta}))u(p_1,s_1) ,\\
			\mathcal{M}^{\pi}_d=
			&\frac{-16g_{\Delta^{\ast} Na_0}g_{\Delta^{\ast}N\pi}g_{\pi\Lambda\Sigma}}{(m_{\Lambda}+m_{\Sigma})m_{\pi}m_{a_0}}f_{\pi}(k_1)f^2_{\Delta^{\ast}}(q)\bar{u}(p_5,s_5)(p_{a_0}^{\mu})\\
			&\Theta_{\mu\nu}(Z2) G^{\nu\rho}_{\Delta^{\ast}}\Theta_{\rho\sigma}(Z2)(\    p_{\pi}^{\sigma})u(p_2,s_2)G_{\pi}(p_{\pi})\\
			&\bar{u}(p_3,s_3)\gamma_5(\slashed{p}_{\pi})u(p_1,s_1) .\\
		\end{split}
	\end{equation}
	The $p_1$, $p_2$, $p_3$, $p_4$, and $p_5$ represent the initial state $\Sigma$ and $p$ particle, as well as the final state $\Lambda$, $a_0$, and $p$ particle, respectively. $k_1$ and $k_2$ correspond to the four-momenta of the exchanged particles in Fig.~\hyperlink{1a}{1(a)} and Fig.~\hyperlink{1b}{1(b)}, respectively, each involving four exchanges. The symbol $q$ has the same meaning as $k_1$ and $k_2$, and applies to both $\Delta(1920)$ and $\Delta(1940)$.
	The differential cross-section and total cross-section for this reaction can be obtained through:
	\begin{equation}
		\begin{split}
			d\sigma=&\dfrac{(2\pi)^4}{4\sqrt{(p_1\cdot p_2)^2-m_p^4}}\left(\dfrac{1}{2}\sum|\mathcal{M}|^2\right)d\Phi_3\\
			=&\dfrac{1}{(2\pi)^4}\dfrac{1}{\sqrt{(p_1\cdot p_2)^2-m_p^4}}\dfrac{|\vec{p}_3||\vec{p}_5^*|}{32\sqrt{s}}\left(\dfrac{1}{2}\sum|\mathcal{M}|^2\right)\\
			&dm_{\Sigma\eta}d\Omega_5^*d\mathrm{cos}\theta_3.
		\end{split}
	\end{equation}
	In this context, $p_1$ and $p_2$ denote the four-momentum of the initial particles $\Sigma$ and $p$, respectively, in the overall center-of-mass frame. Additionally, $\vec{p}_5^*$ represents the three-momentum of the $p$ baryon within the center-of-mass frame of the $a_0p$ pair.
	
	\section{results}
	\begin{figure}[htpb]  
		\centering
		\includegraphics[width=0.45\textwidth]{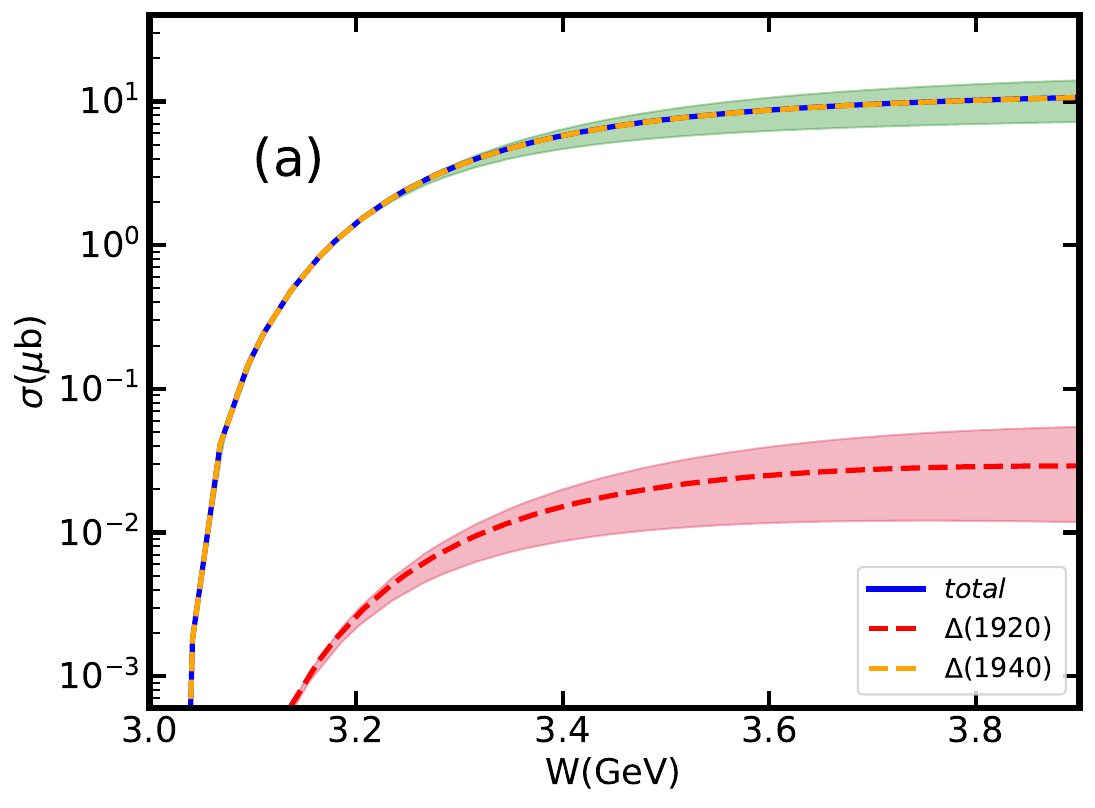}\hypertarget{2a}{}\\
		\includegraphics[width=0.45\textwidth]{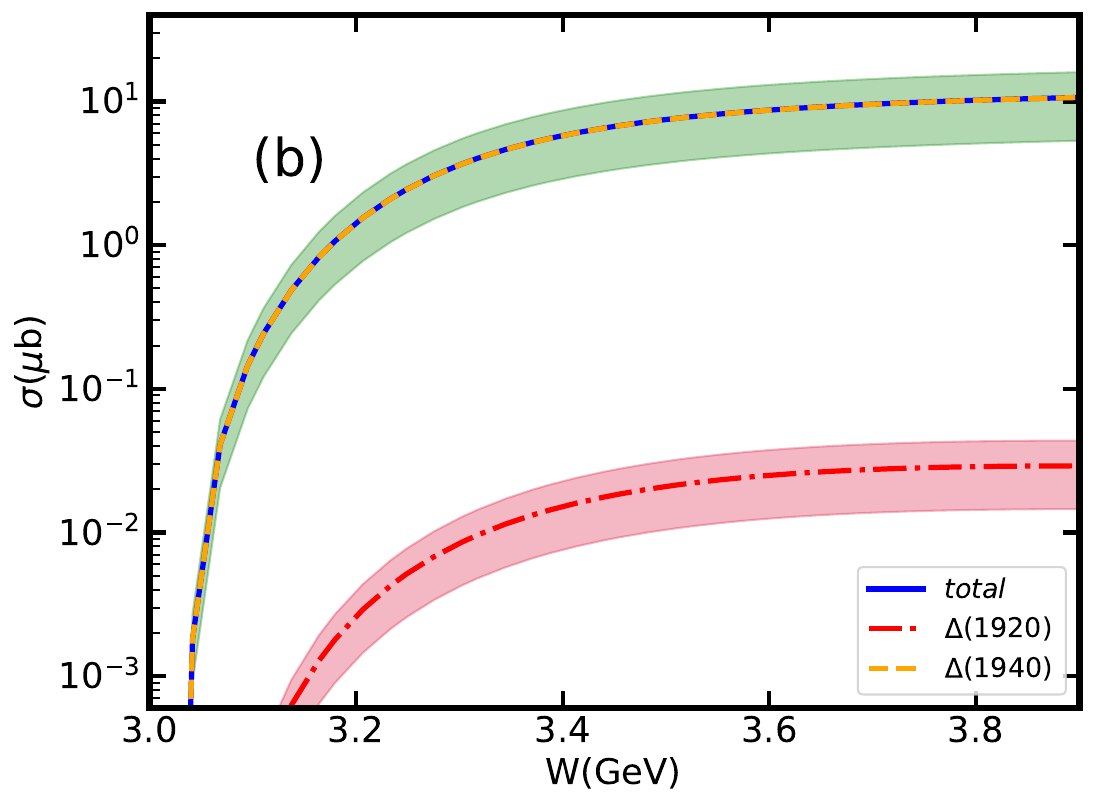}\hypertarget{2b}{}
		\captionsetup{justification=raggedright}
		\caption{The total cross section of the $\Sigma^{+} p\to\Lambda a_0^{+} p$ reaction as a function of the center-of-mass energy W. The blue curve represents the total cross section including all contributions depicted in Fig.~\ref{1}, while the red and orange dashed lines correspond to the contributions from $\Delta(1920)$ and $\Delta(1940)$ respectively. The green and red bands represent the variation ranges of $\Delta(1940)$ and $\Delta(1920)$ respectively. The above figure shows the variation of the cutoff parameter $\Lambda_{\Delta(1940)}$ between 1.2~GeV and 1.8~GeV, and the cutoff parameter $\Lambda_{\Delta(1920)}$ between 1.2~GeV and 1.8~GeV. The below figure illustrates the variation of the branching ratio for the $\Delta(1920)\to a_0p$ decay between 2\% and 6\%, and the branching ratio for the $\Delta(1940)\to a_0 p$ decay between 1\% and 3\%.}
		\label{2}
	\end{figure}
\begin{figure}[htbp]
		\centering
		\includegraphics[width=0.45\textwidth]{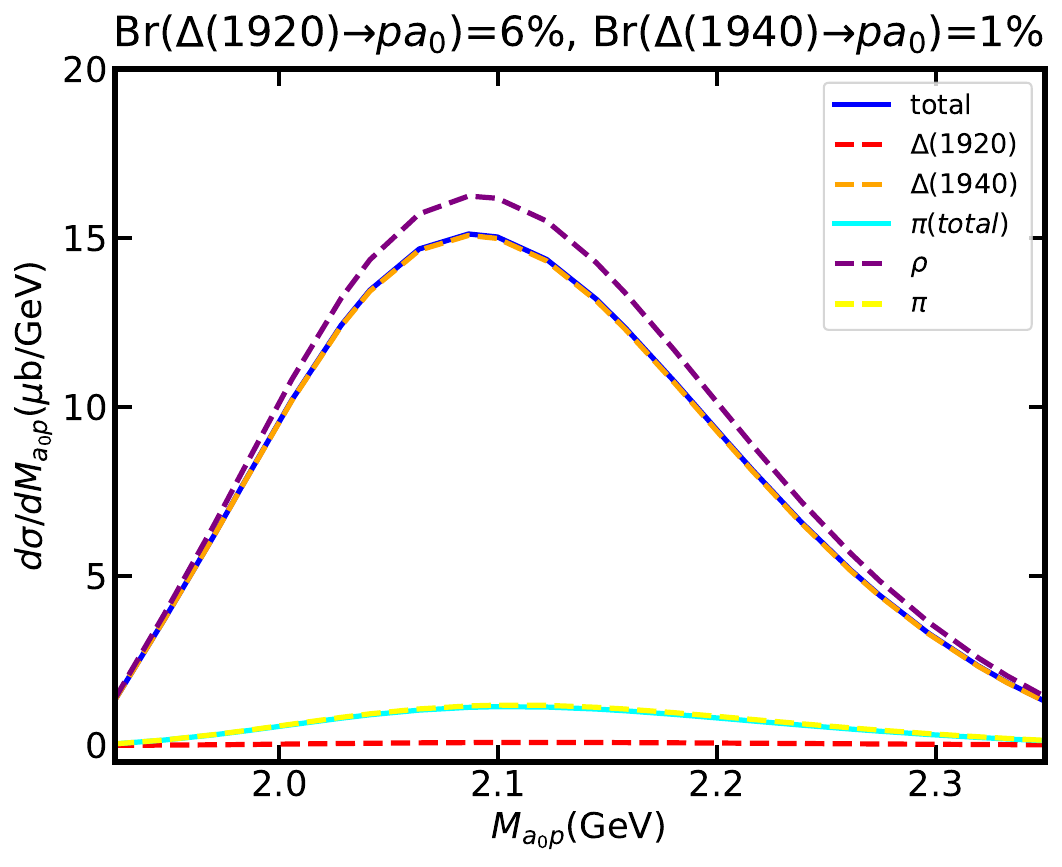}
		\captionsetup{justification=raggedright}
		\caption{At center of mass energy $W=3.5$~GeV, the invariant mass distribution of final $a_0 p$ pair in the worst case with Br$(\Delta(1940)\to pa_0)=$1\% and Br$(\Delta(1920)\to pa_0)=$6\%.}
		\label{10}
	\end{figure}

	In this section, we engage in a series of discussions on the theoretical results of the total and differential cross-sections for the $\Sigma^{+} p \to \Lambda a_0^{+} p$ reaction, which were calculated using the model from the previous section. Firstly, we consider the impact of the branching ratio Br$(\Delta^{\ast} \to a_0 p)$ on the total cross-section by fixing the cutoff parameter $\Lambda_{\Delta^{\ast}}$ at 1.5~GeV. Prior to this, we need to neglect the off-shell effects and set Z = -0.5. In Fig.~\ref{2}, we plot the total cross-section from the reaction threshold to 3.9~GeV, taking into account the contributions of both $\Delta(1920)$ and $\Delta(1940)$ resonances. In our model, $\Delta^{\ast}$ is dependent on the model parameters. The two colored bands in Fig.~\hyperlink{2a}{2(a)} represent the range of variation of the cutoff parameter $\Lambda_{\Delta^{\ast}}$ from 1.2~GeV to 1.8~GeV. From Fig.~\hyperlink{2a}{2(a)}, it can be seen that the variation of the cutoff parameter has a more sensitive effect on the contribution of $\Delta(1920)$ compared to that of $\Delta(1940)$. However, the cutoff parameter is generally regarded as a free parameter and thus requires more experimental data to determine it. Additionally, we have also considered the influence of the branching ratio of the $\Delta^{\ast} \to a_0 p$ decay on the cross-section, as shown in Fig.~\hyperlink{2b}{2(b)}. Regardless of how the value of the branching ratio changes, it cannot alter the dominant role played by the contribution of $\Delta(1940)$. Even when Br$(\Delta(1920) \to a_0 p) = 6\%$ takes its maximum value and Br$(\Delta(1940) \to a_0 p) = 1\%$ takes its minimum value, as shown in Fig.~\ref{10}, the contribution of $\Delta(1940)$ in this reaction is still significantly greater than that of $\Delta(1920)$. Neither changing the cutoff parameter nor the value of the branching ratio can alter the phenomenon that the contribution of $\Delta(1940)$ is more significant than that of $\Delta(1920)$. On the one hand, this is due to the relatively large couplings of both $\Delta(1940)N\pi$ and $\Delta(1940)N\rho$. On the other hand, the lack of coupling between $\Delta(1920)$ and the $N\rho$ channel results in the absence of the reaction process shown in Fig.~\hyperlink{1a}{1(a)}, which in turn weakens the contribution of $\Delta(1920)$. Even when focusing solely on $\pi$ exchange, the contribution of $\Delta(1940)$ is still significantly greater than that of $\Delta(1920)$. Given the dominance of $\Delta(1940)$ in the region close to the reaction threshold, this reaction is considered an ideal research platform for further exploring the unique properties of the $\Delta(1940)$ resonance. Through this approach, we can more accurately analyze and understand the intrinsic characteristics of this particle.
	\begin{figure}[htpb]
		\centering
		\includegraphics[width=0.45\textwidth]{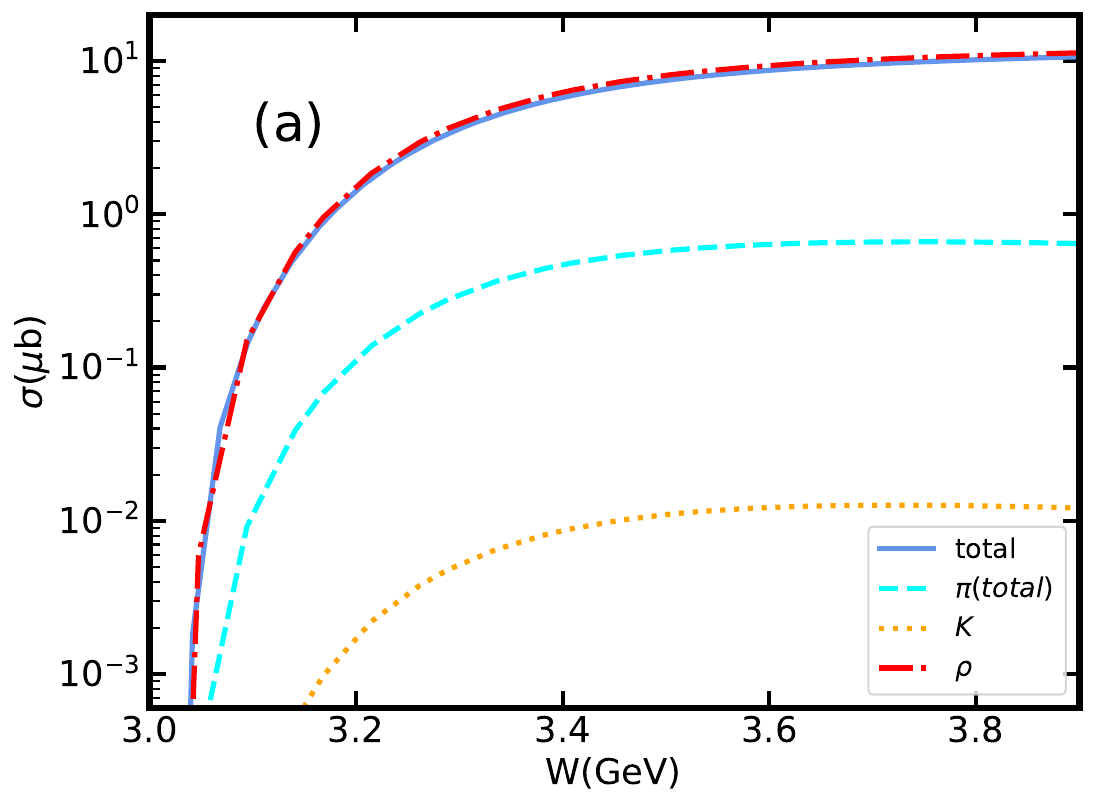}\hypertarget{3a}{}\\
		\includegraphics[width=0.45\textwidth]{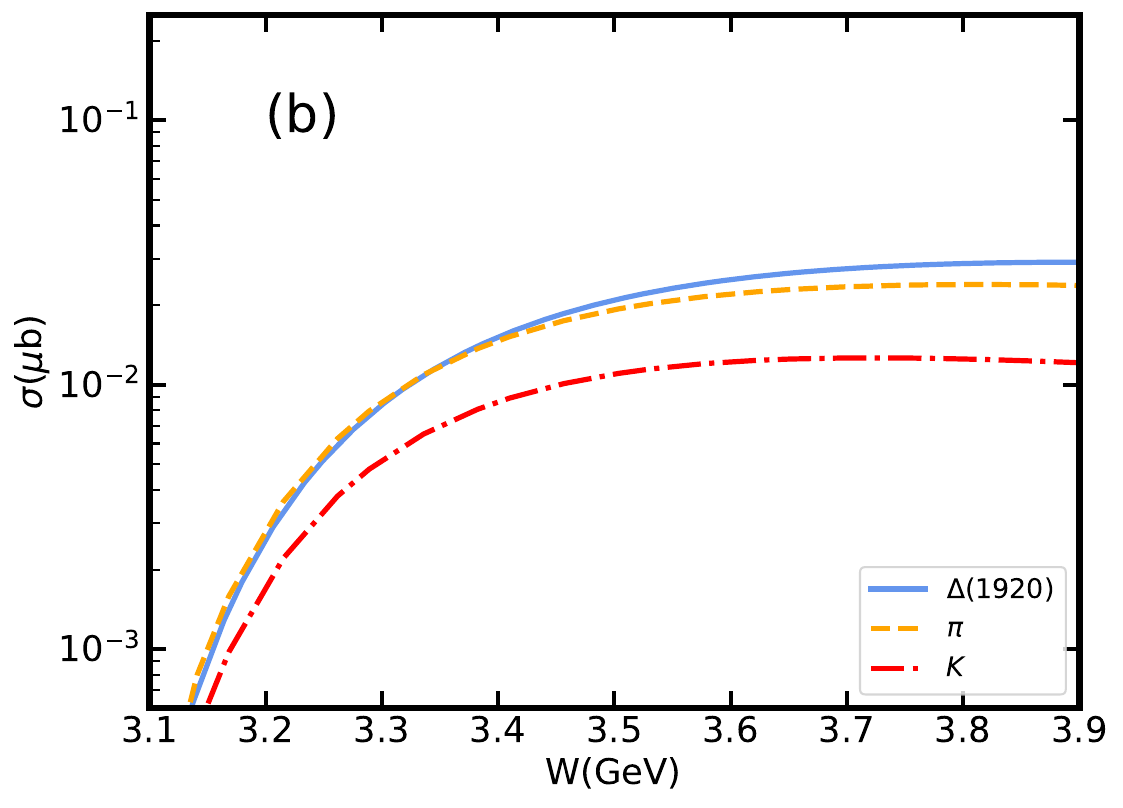}\hypertarget{3b}{}\\
		\includegraphics[width=0.45\textwidth]{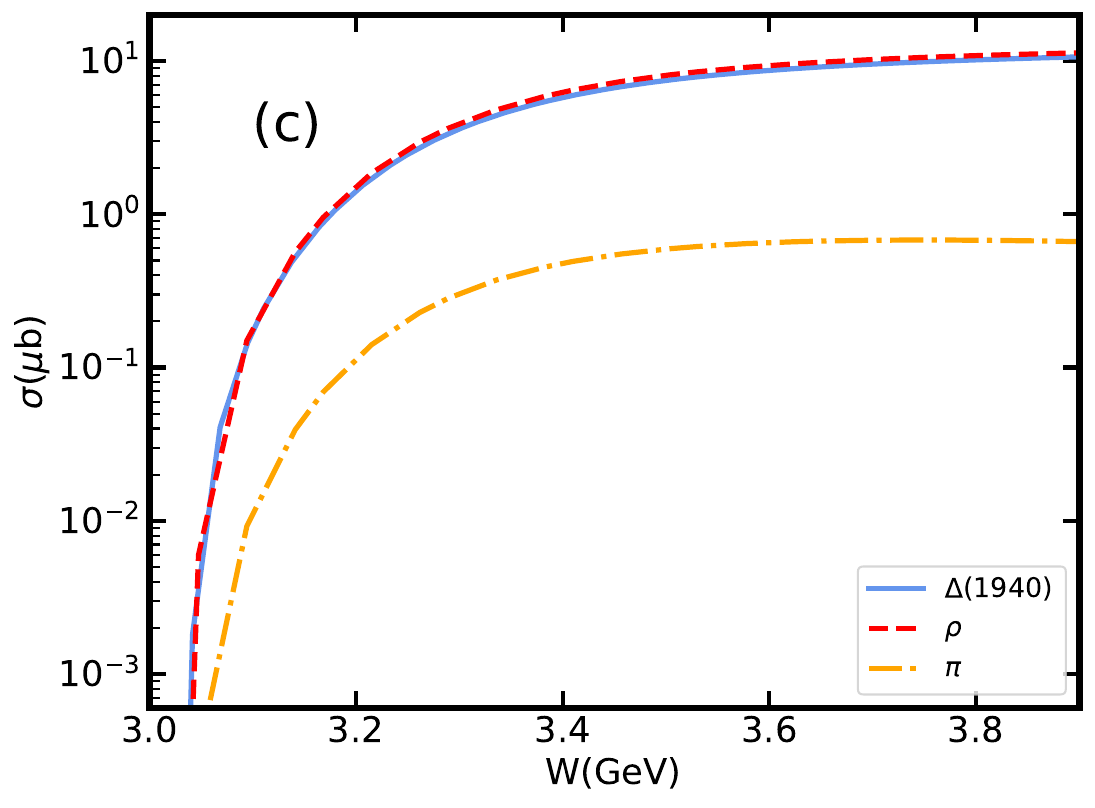}\hypertarget{3c}{}
		\captionsetup{justification=raggedright}
		\caption{The relationship between the cross sections obtained through meson exchanges involving $\pi$ and $K$, as well as $\pi$ and $\rho$, and the center-of-mass energy W.}
		\label{3}
	\end{figure}
	Due to the uncertainty of the form factors, it is necessary to consider their impact on the cross-sections here. In Figs.~\hyperlink{3a}{3(a)}-~\hyperlink{3c}{3(c)}, we present the cross-sections for the $\Delta(1920)$ via meson $\pi$ and $K$ exchanges, the cross-sections for the $\Delta(1940)$ via meson $\pi$ and $\rho$ exchanges, and a comparison of the contributions of individual meson exchanges and the total cross-sections. From Fig.~\hyperlink{3a}{3(a)} and Fig.~\hyperlink{3c}{3(c)}, it can be seen that the contribution of $\Delta(1940)$ dominates, with the $\rho$ meson contributing the most significantly, followed by the $\pi$ meson. Even when focusing solely on the $\pi$ meson, the contribution of $\Delta(1940)$ is still larger than that of $\Delta(1920)$. For $\Delta(1920)$, the contribution of the $\pi$ meson is relatively large, followed by the $K$ meson. Additionally, both $\Delta(1920)$ and $\Delta(1940)$ involve $\pi$ meson exchanges, but there are significant numerical differences. This is due to the $\mathcal{D}$-wave nature of the $\Delta(1940)N\pi$ coupling and the large threshold momentum of this reaction, which, compared to $\Delta(1920)$, enhance the contribution of $\Delta(1940)$.
	
	\begin{figure}[htpb]
		\centering
		\includegraphics[width=0.45\textwidth]{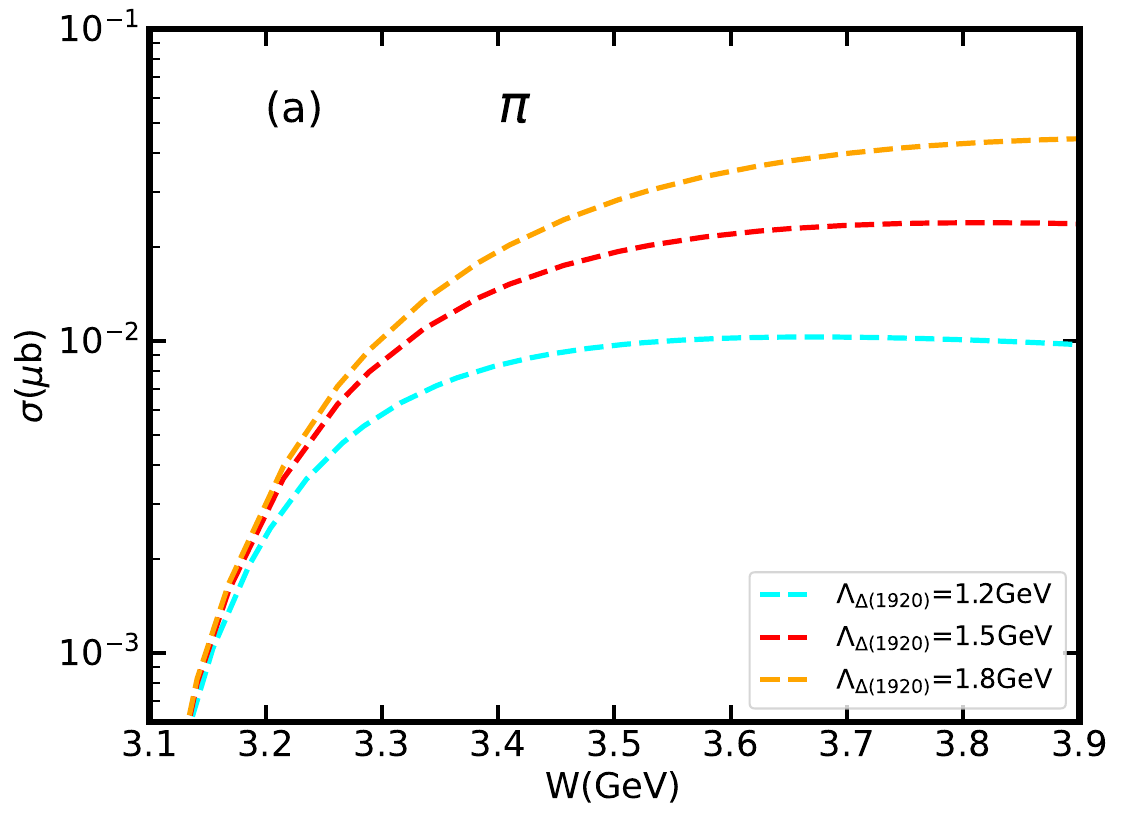}\hypertarget{4a}{}\\
		\includegraphics[width=0.45\textwidth]{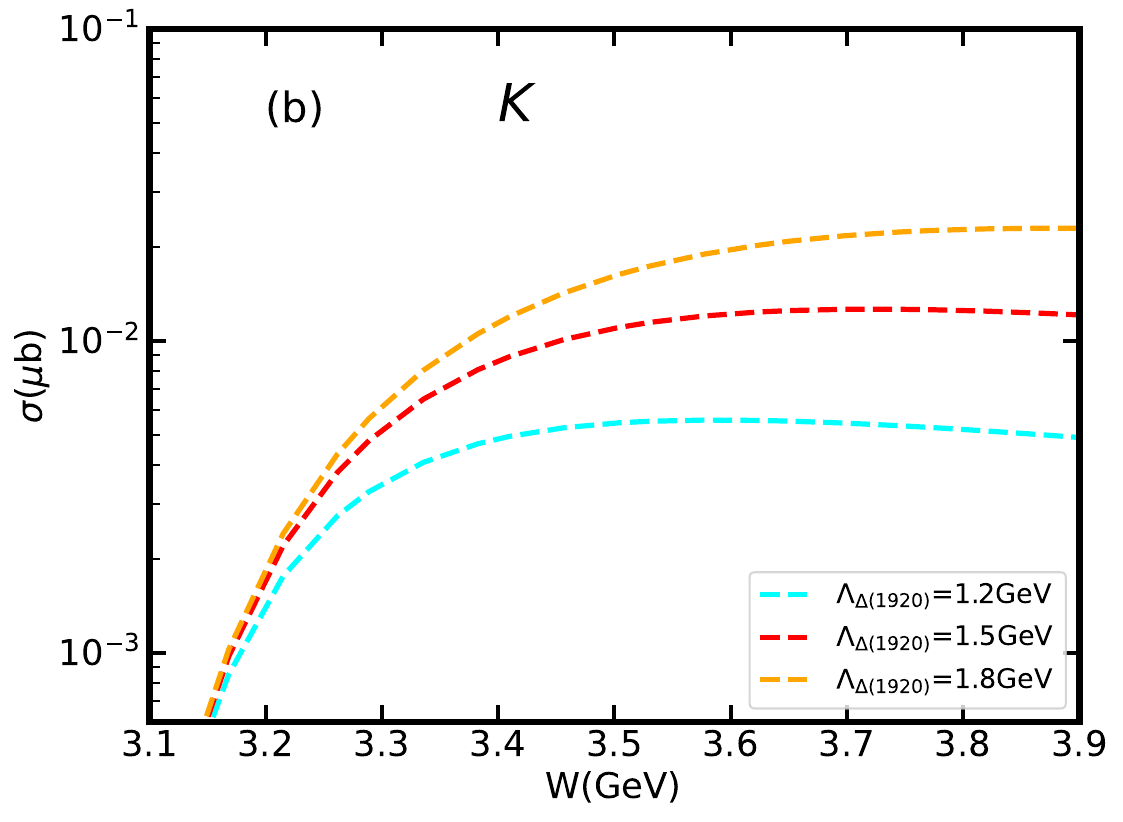}\hypertarget{4b}{}\\
		\includegraphics[width=0.45\textwidth]{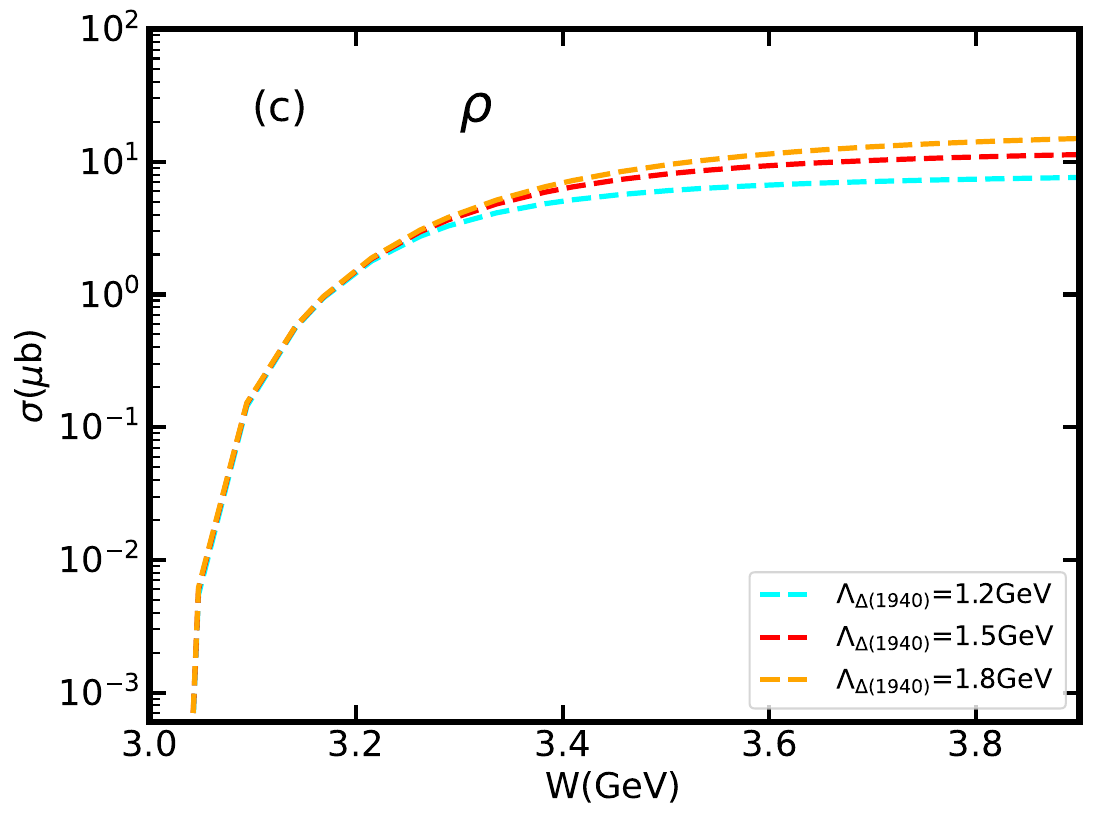}\hypertarget{4c}{}\\
		\includegraphics[width=0.45\textwidth]{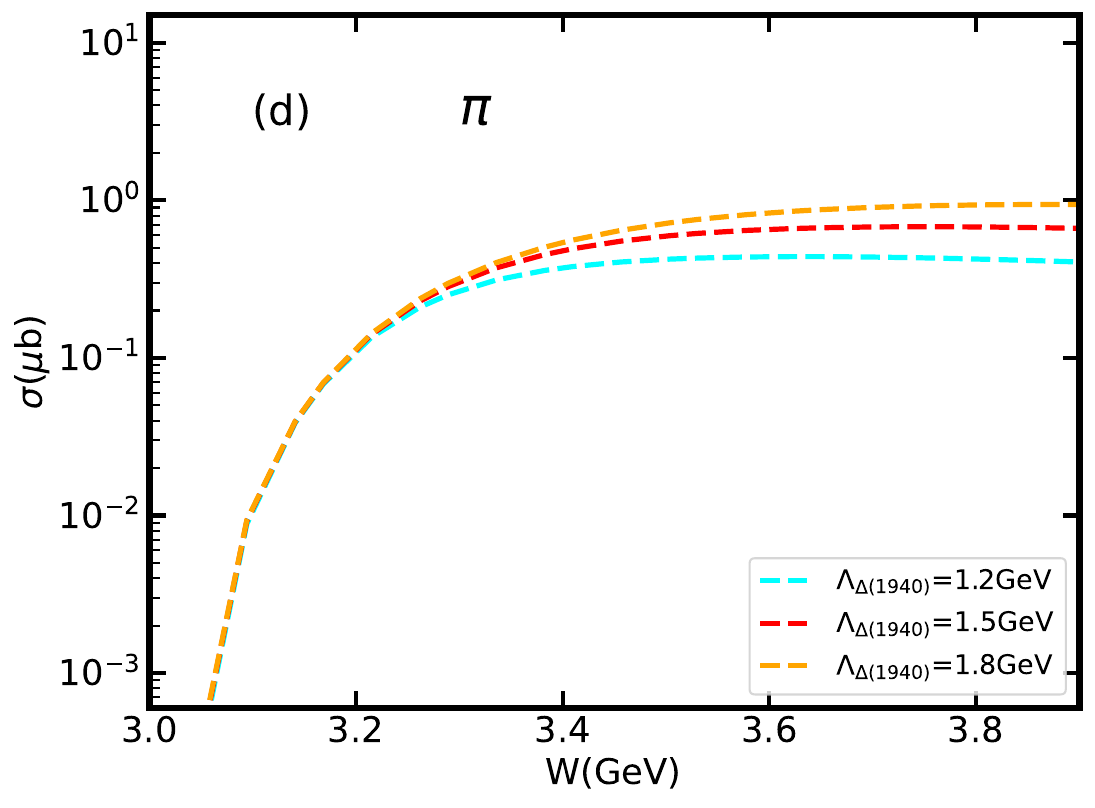}\hypertarget{4d}{}\\
		\captionsetup{justification=raggedright}
		\caption{The relationship between the cross sections of various meson exchanges and the center-of-mass energy W, with varying values of the cutoff parameter $\Lambda_{\Delta{\ast}}$.}
		\label{4}
	\end{figure}
	Next, we consider the dependence of the cross-section on the cutoff parameter $\Lambda_{\Delta^{*}}$ introduced by the form factor. In Fig.~\ref{4}, we present the trends in the cross-section for three cutoff parameters, $\Lambda_{\Delta^{*}}$ = 1.2, 1.5, and 1.8~GeV. As the cutoff parameter increases, there is a certain degree of increase in the contributions of both $\Delta(1920)$ and $\Delta(1940)$. Regardless of whether it is near the threshold or at higher energies, the $\pi^{+}$ meson exchange is more sensitive to changes in the cutoff parameter $\Lambda_{\Delta^{*}}$.
	\begin{figure}[htpb]
		\centering
		\includegraphics[width=0.45\textwidth]{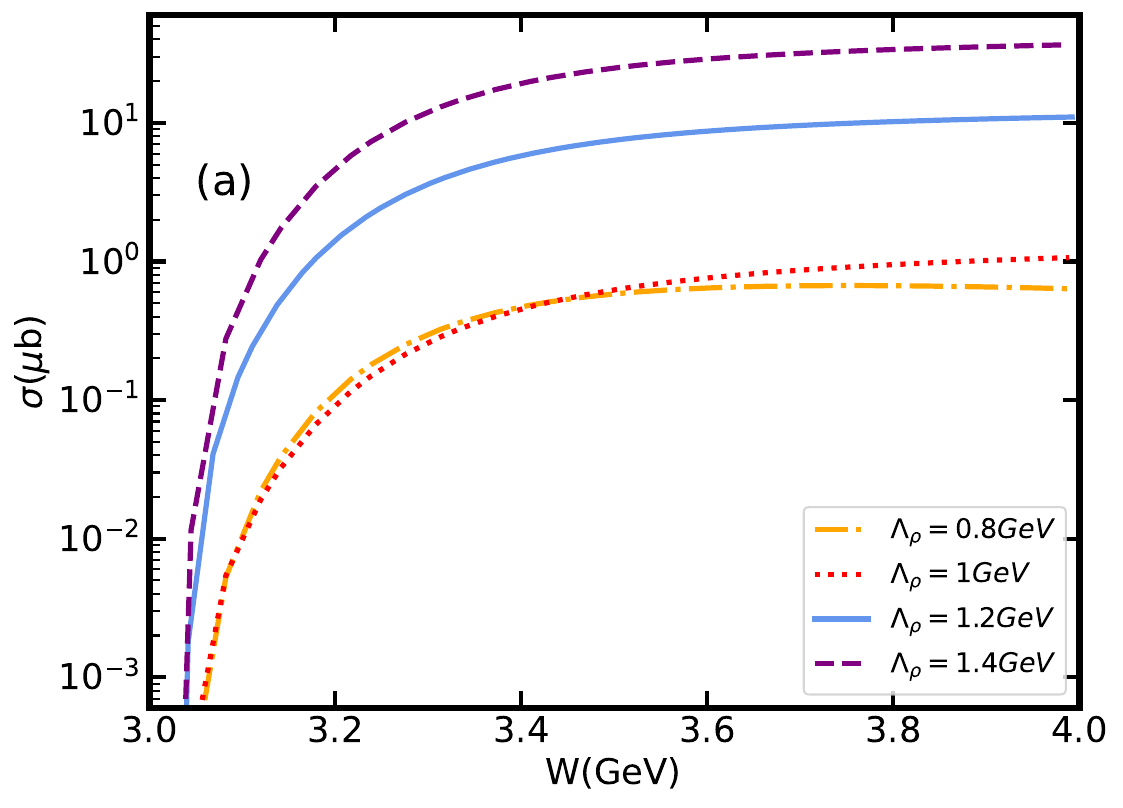}
		\includegraphics[width=0.45\textwidth]{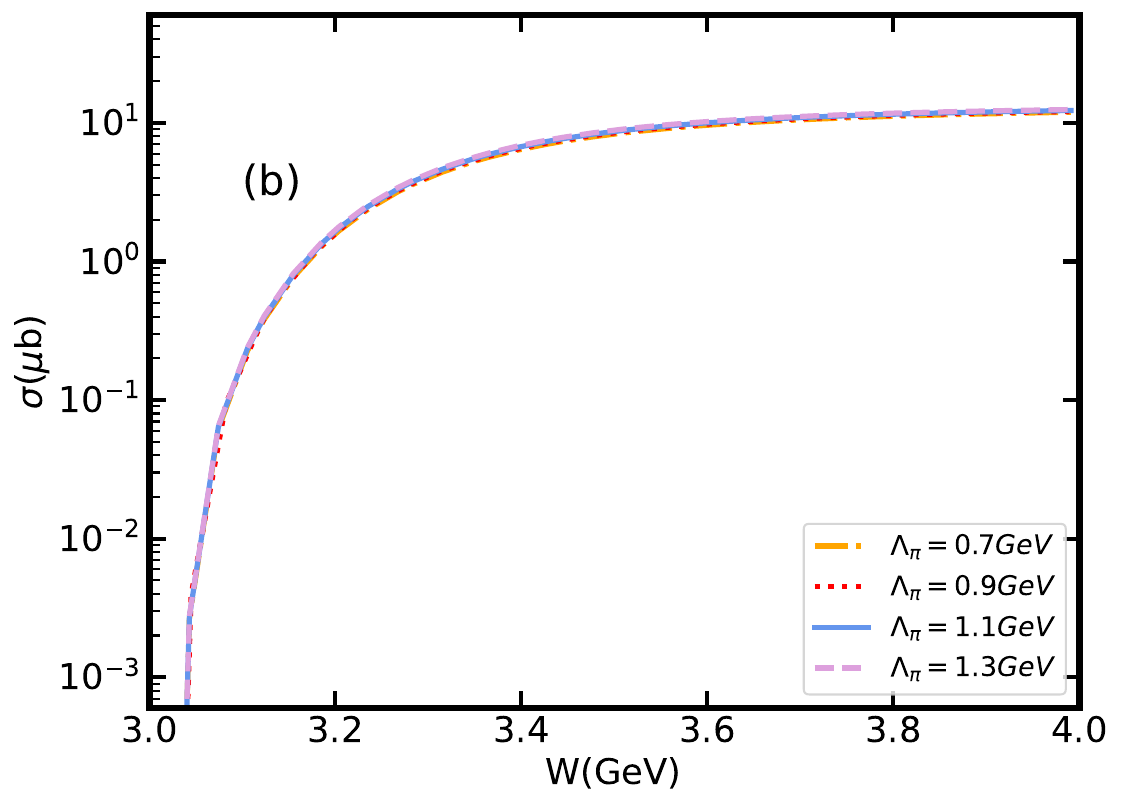}
		\captionsetup{justification=raggedright}
		\caption{The influence of variations in the cutoff parameters for $\pi$ and $\rho$ mesons on the cross section.}
		\label{5}
	\end{figure}
   Here, we consider the impact of $\rho$ and $\pi$ mesons on the contribution of $\Delta(1940)$ by varying the cutoff parameters. In Fig.~\hyperlink{5a}{5(a)}, we present the changes in the cross-section contribution of $\Delta(1940)$ observed at cutoff parameters $\Lambda_{\rho}$ = 0.8, 1.0, 1.2, and 1.4~GeV. In Fig.~\hyperlink{5b}{5(b)}, we show the variation in the $\Delta(1940)$ cross-section at cutoff parameters $\Lambda_{\pi}$ = 0.7, 0.9, 1.1, and 1.3~GeV. The results indicate that the overall dependence of the $\Delta(1940)$ contribution on $\Lambda_{\rho}$ is stronger compared to $\Lambda_{\pi}$. The contribution of $\Delta(1940)$ is more sensitive to increases in the $\Lambda_{\rho}$ cutoff parameter, while changes in the $\Lambda_{\pi}$ cutoff parameter have a negligible impact on the $\Delta(1940)$ contribution.
	
	\begin{figure}[htpb]
		\centering
		\includegraphics[width=0.45\textwidth]{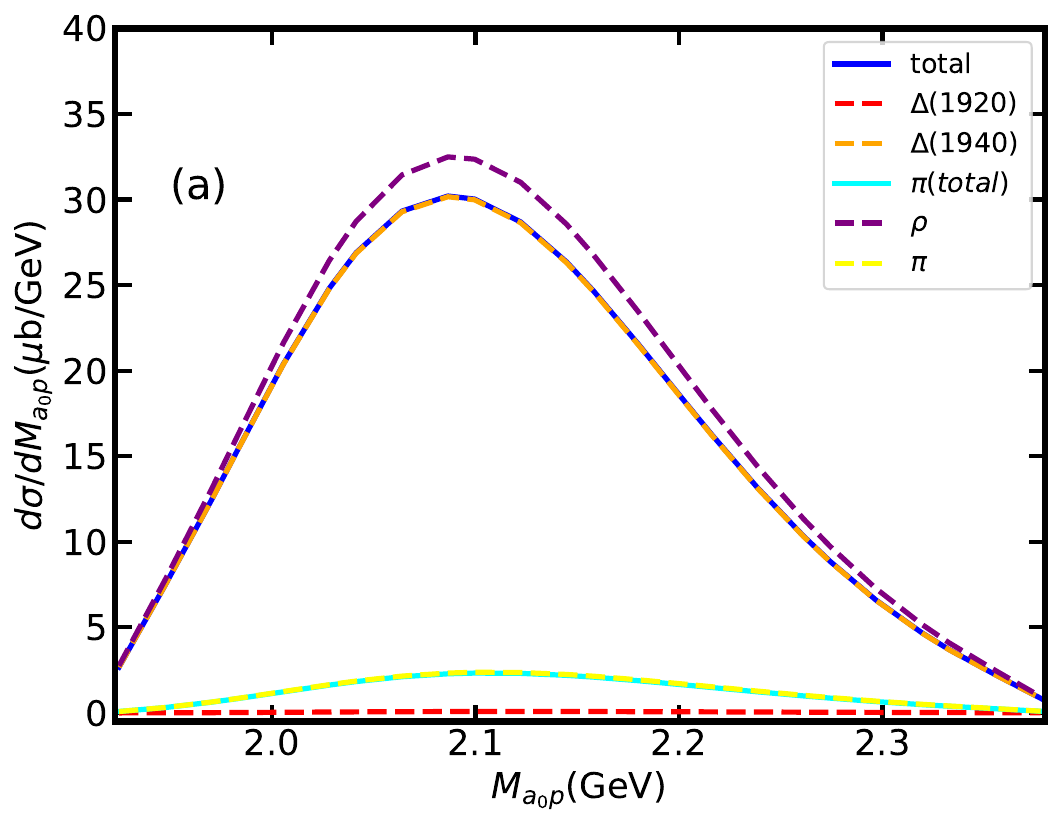}\hypertarget{6a}{}\\
		\includegraphics[width=0.45\textwidth]{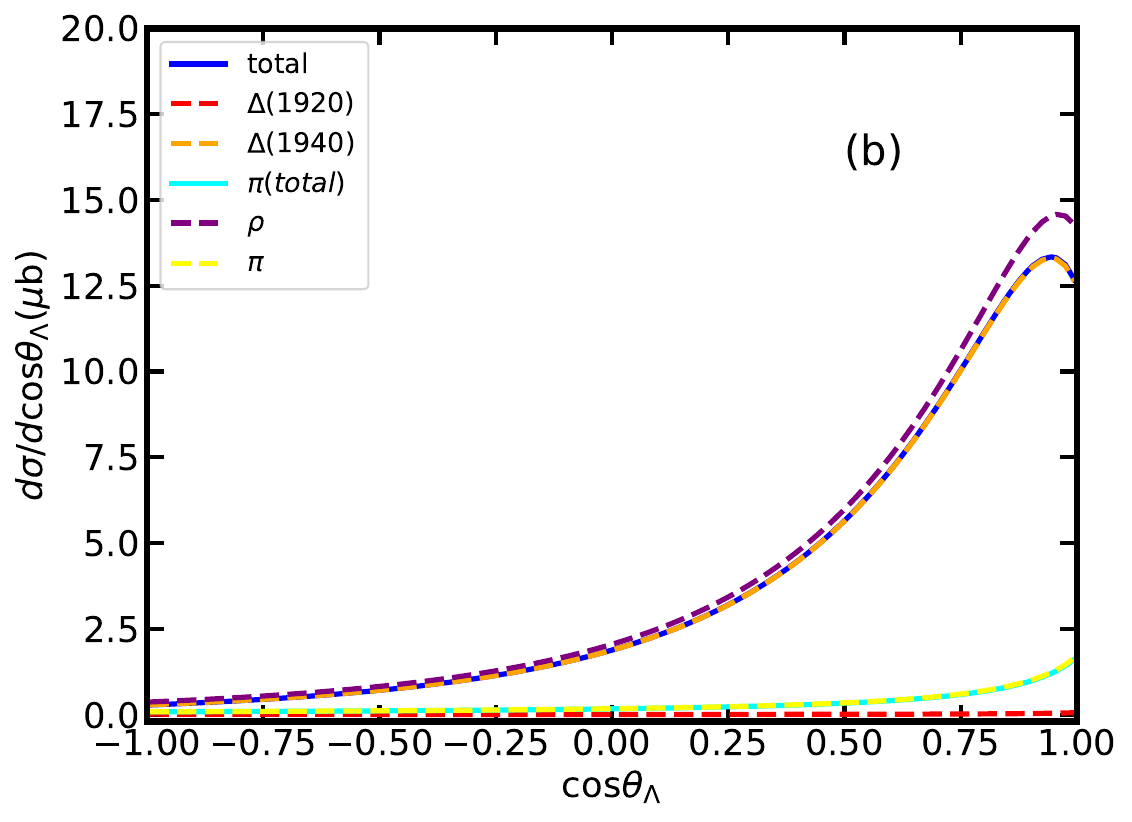}\hypertarget{6b}{}
		\captionsetup{justification=raggedright}
		\caption{At a center-of-mass energy $W = 3.5$~GeV, (a) the invariant mass distribution of the final state $a_0 p$ pair; (b) the angular distribution of $\Lambda$. The blue curve in figures represents the total cross section including all contributions depicted in Fig.~\ref{1}. The red and yellow dashed lines correspond to the contributions from $\Delta(1920)$ and $\Delta(1940)$ respectively. The fluorescent blue, purple, and yellow dashed lines represent the contributions from $\pi$ meson exchange within $\Delta(1920)$ and $\Delta(1940)$, as well as the combined contributions from $\rho$ meson and $\pi$ meson exchange within $\Delta(1940)$.}
		\label{6}
	\end{figure}
	In addition to the total cross-section, we have also investigated the differential cross-section for the reaction $\Sigma^{+} p \to \Lambda a_{0}^{+} p$ at a center-of-mass energy $W = 3.5$~GeV. Fig.~\hyperlink{6a}{6(a)} shows the contribution to the $a_{0}p$ invariant mass distribution. It can be seen from Fig.~\hyperlink{6a}{6(a)} that $\Delta(1940)$ is the main contributor, with a distinct peak in its contribution. Due to the large resonance width of $\Delta(1940)$, the curve shape is relatively flat. In Fig.~\hyperlink{6b}{6(b)}, the total contribution from $\rho$ exchange is evident as a forward enhancement in the angular distribution of $\Lambda$, and it significantly affects the shape of the total contribution. In contrast, the contribution of $\pi$ exchange via the $t$-channel is relatively small in the angular distribution. Compared to the contribution of $\Delta(1940)$, the contribution of $\Delta(1920)$ in the angular distribution of $\Lambda$ is not significant.
	
	\begin{figure}[htpb]
		\centering
		\includegraphics[width=0.45\textwidth]{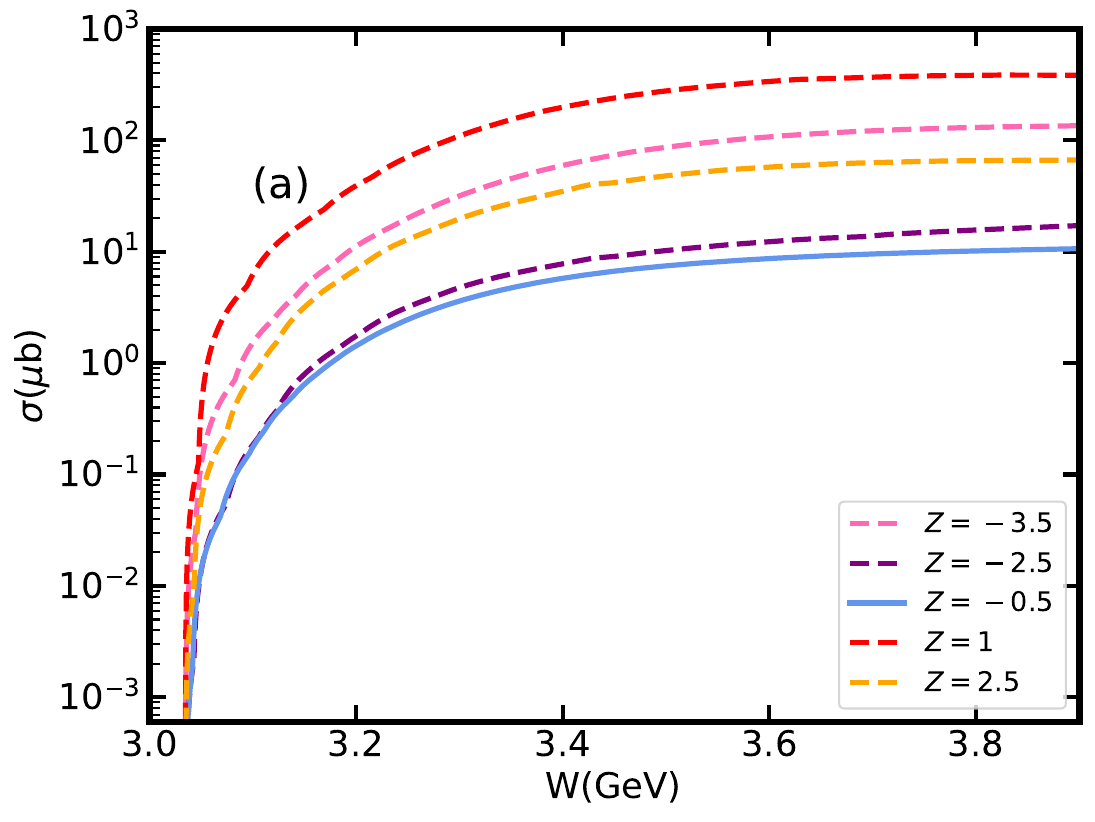}\hypertarget{7a}{}\\
		\includegraphics[width=0.45\textwidth]{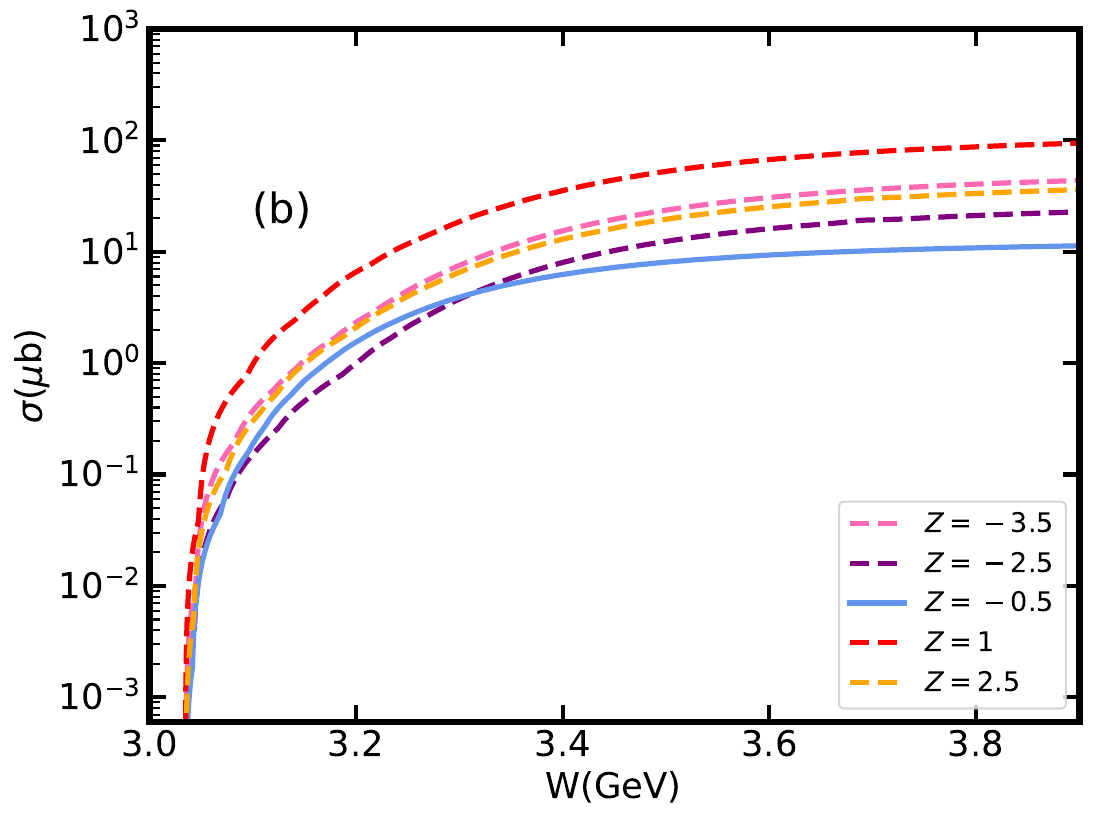}\hypertarget{7b}{}\\
		\includegraphics[width=0.45\textwidth]{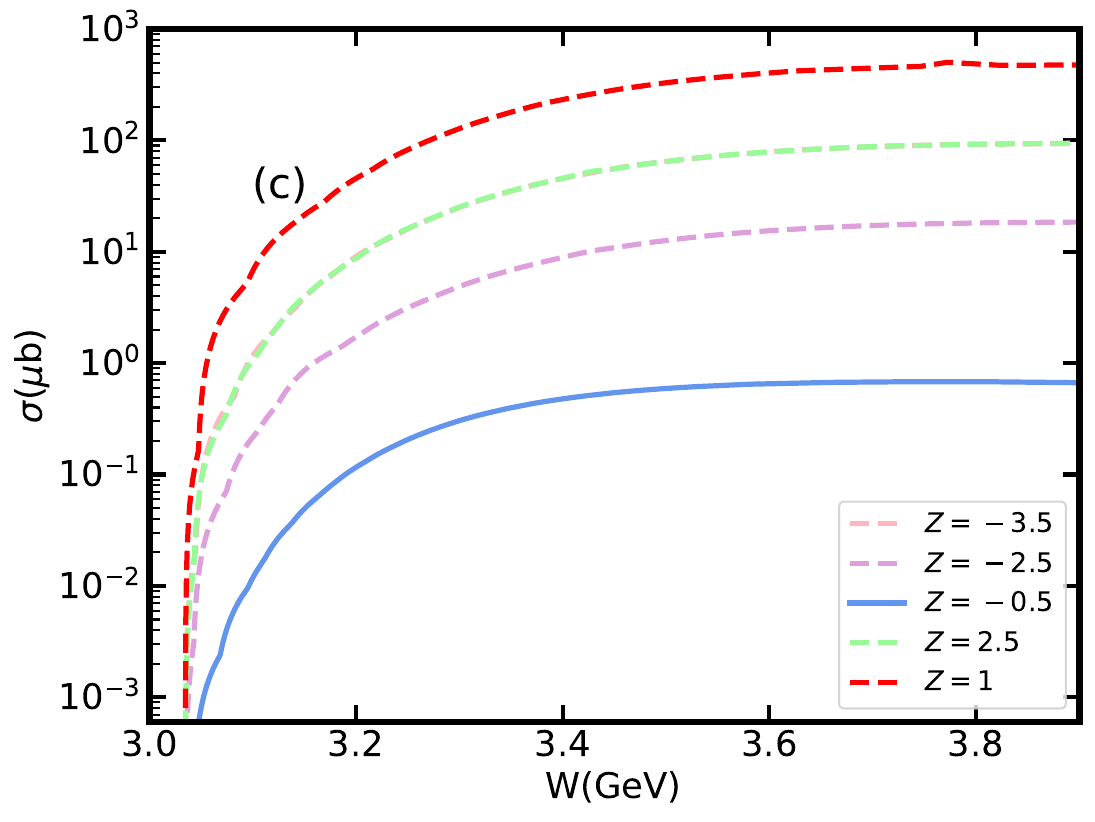}\hypertarget{7c}{}
		\captionsetup{justification=raggedright}
		\caption{Considering the cross-section contributions with off-shell effects, where (a), (b) and (c) represent the total cross-section, the contribution from $\rho$ meson exchange, and the contribution from $\pi$ meson exchange, respectively.}
		\label{7}
	\end{figure}
	
\begin{figure}[htpb]
		\centering
		\includegraphics[width=0.45\textwidth]{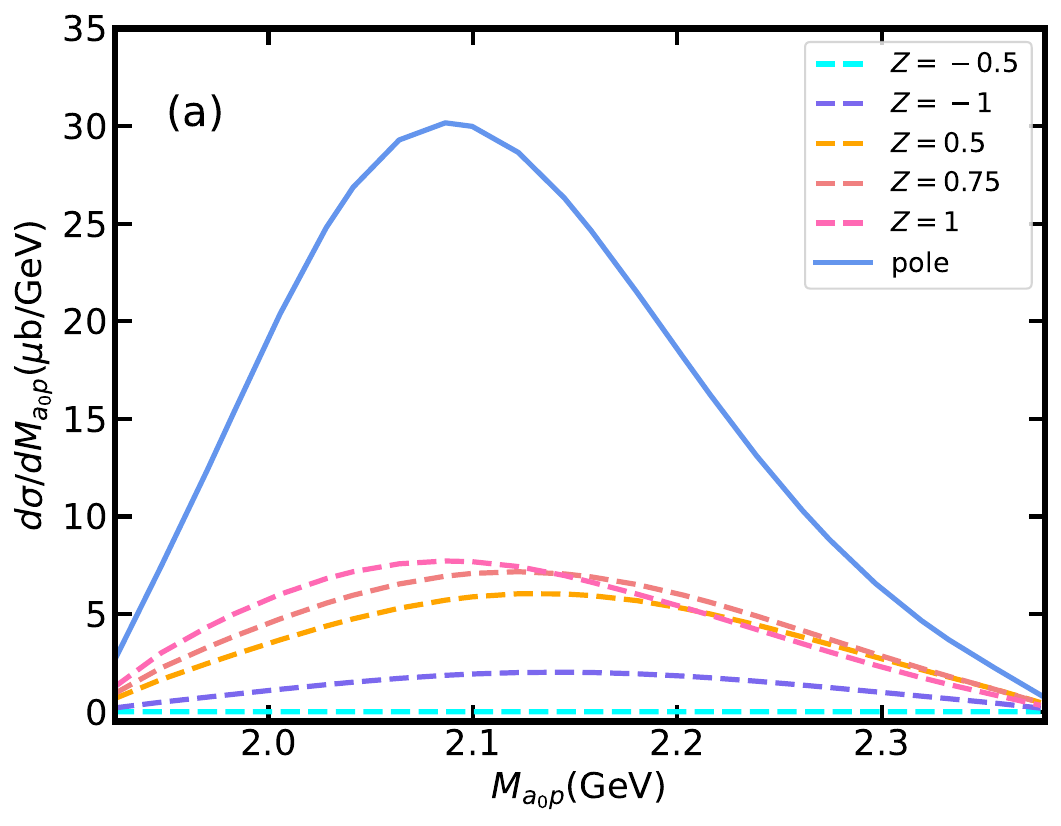}\hypertarget{8a}{}\\
		\includegraphics[width=0.45\textwidth]{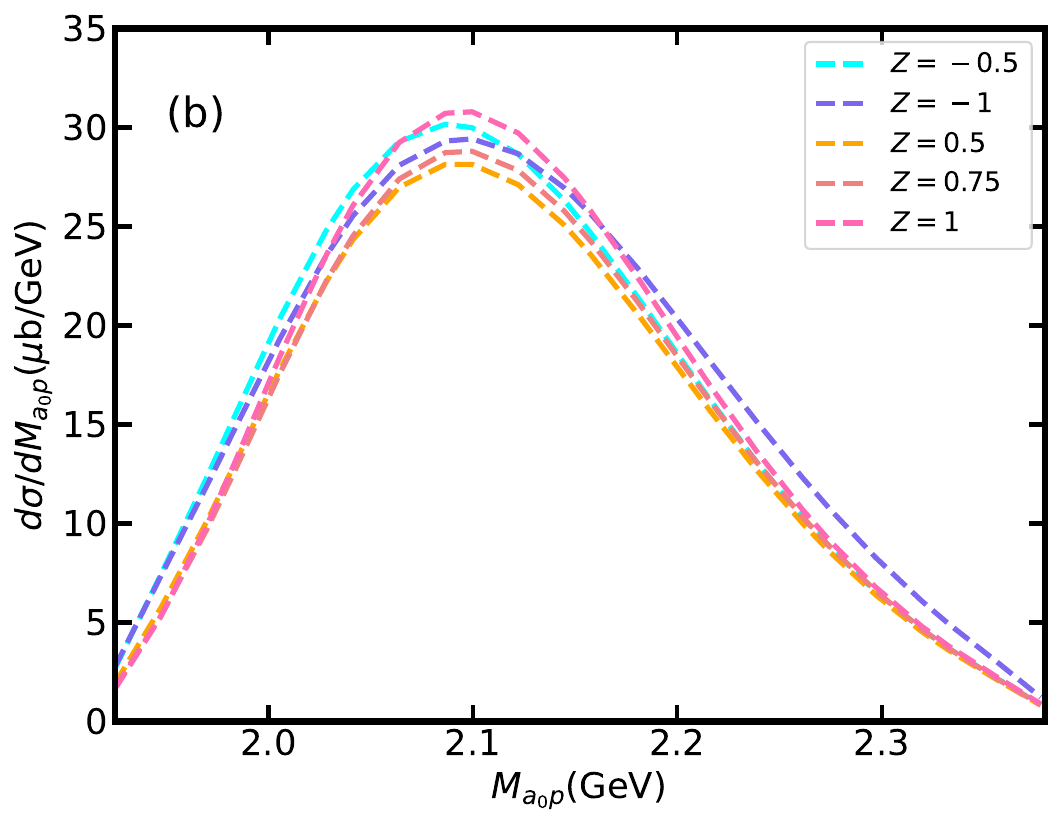}\hypertarget{8a}{}
		\captionsetup{justification=raggedright}
		\caption{At center of mass energy $W=3.5$~GeV, the invariant mass distribution of final $a_0 p$ pair with taking into account of the off-shell effects.}
		\label{8}
	\end{figure}

	We analyze how the off-shell parameters affect the cross-section, with particular attention to their impact on the contribution to the $\Delta(1940)$ cross-section. Fig.~\ref{8} visually demonstrate the dynamic changes in the cross-section under different off-shell parameter conditions. Specifically, Fig.~\hyperlink{8a}{8(a)} depicts the cross-section variation contributed solely by $\Delta(1940)$, while Fig.~\hyperlink{8b}{8(b)} and Fig.~\hyperlink{8c}{2(c)} represent the contributions to the cross-section from the $\rho$-meson and $\pi$-meson exchange processes, respectively.
	
	After analyzing the graphical data, we can clearly observe the following conclusion: Compared to the contribution of the $\pi$ meson exchange process to the cross-section, the off-shell parameters have a relatively smaller impact on the cross-section contribution of the $\Delta(1940)$ and $\rho$ meson exchange processes. This clearly indicates that this process exhibits lower sensitivity to variations in the off-shell parameters. Conversely, the contribution of the cross-section in the $\pi$ meson exchange process is notably influenced by changes in the off-shell parameters, indicating a higher sensitivity to variations in such parameters.This difference may stem from the fact that the effective Lagrangian governing the $\Delta(1940)N\rho$ vertex does not directly contain off-shell parameters, causing the amplitude of the $\rho$-meson exchange to be primarily controlled by the off-shell parameters introduced at another vertex, $\Delta(1940)Na_0$, while both vertices related to the $\pi$-meson, $\Delta(1940)N\pi$ and $\Delta(1940)Na_0$, contain off-shell parameters. Therefore, despite variations in the off-shell parameters, the cross-section contribution from the $\rho$-meson exchange exhibits relatively stable characteristics and does not respond significantly to minor changes in the off-shell parameters. This finding enhances our understanding of the cross-section contribution of $\Delta(1940)$ and provides a reference for exploring off-shell effects in other resonant states and interactions.
	
	Then, we focus on studying the impact of the off-shell effect on the $a_0p$ invariant mass spectrum, closely following up on previous literature. We conduct a detailed analysis of the amplitude of the $\Delta(1940)$ resonant state, dividing it into polar and non-polar components. The polar component is independent of the off-shell parameter Z, while the influence of the off-shell parameter is confined to the non-polar component, and the non-polar contribution does not always exceed the polar contribution. Fig.~\hyperlink{8a}{8(a)} shows the separation of polar and non-polar contributions, while Fig.~\hyperlink{8b}{8(b)} displays the total contribution from $\Delta(1940)$. The study finds that although the off-shell parameter affects the overall shape of the $a_0p$ invariant mass distribution, the polar component stably reflects the structure of $\Delta(1940)$ and is not influenced by changes in the off-shell parameter. Therefore, the off-shell parameter has a insignificant impact on the peak position and curve shape in the $a_0p$ invariant mass distribution.
	\begin{figure*}[htpb]
		\centering
		\includegraphics[width=0.4\textwidth]{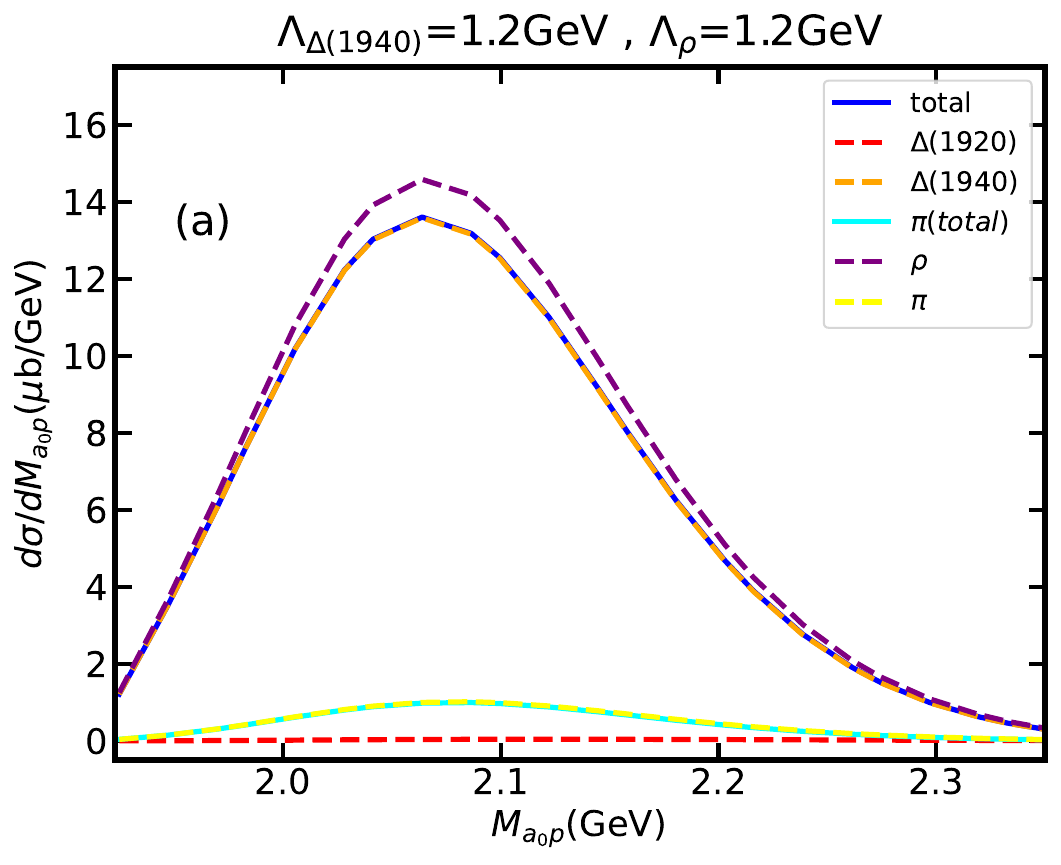}\hypertarget{a}{}
		\includegraphics[width=0.4\textwidth]{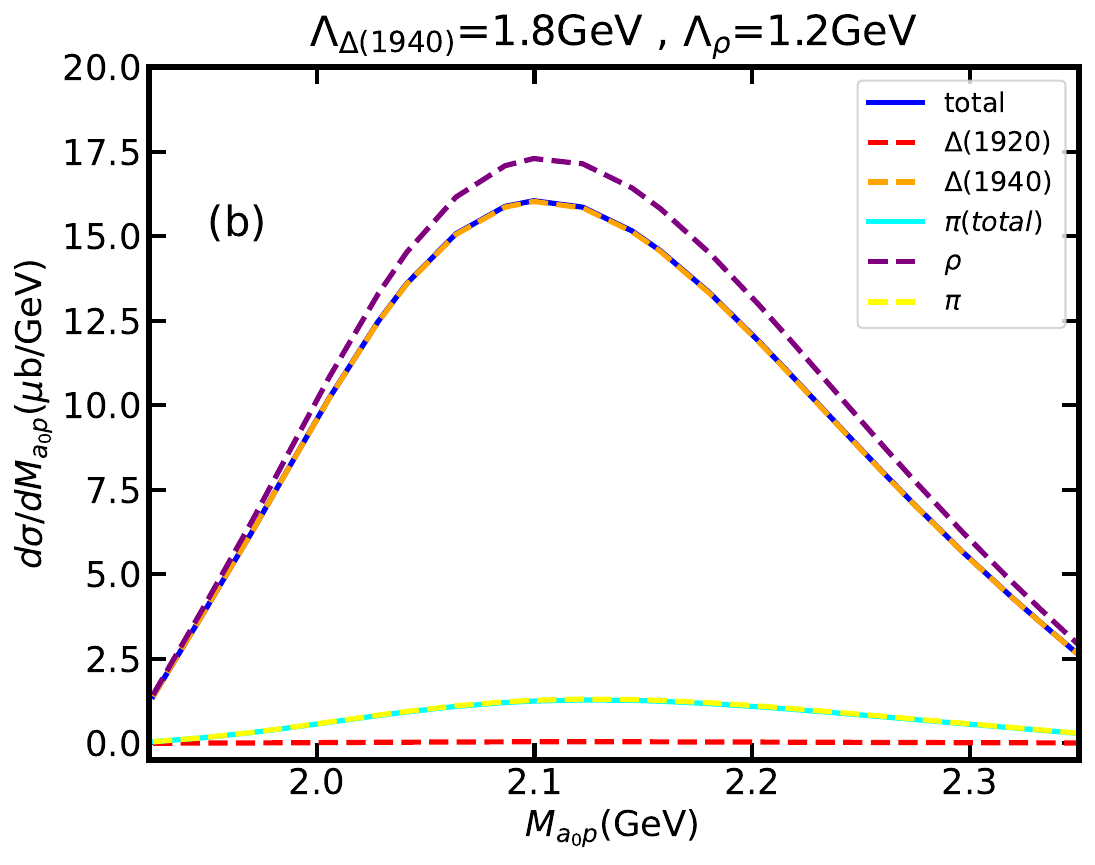}\\
		\includegraphics[width=0.4\textwidth]{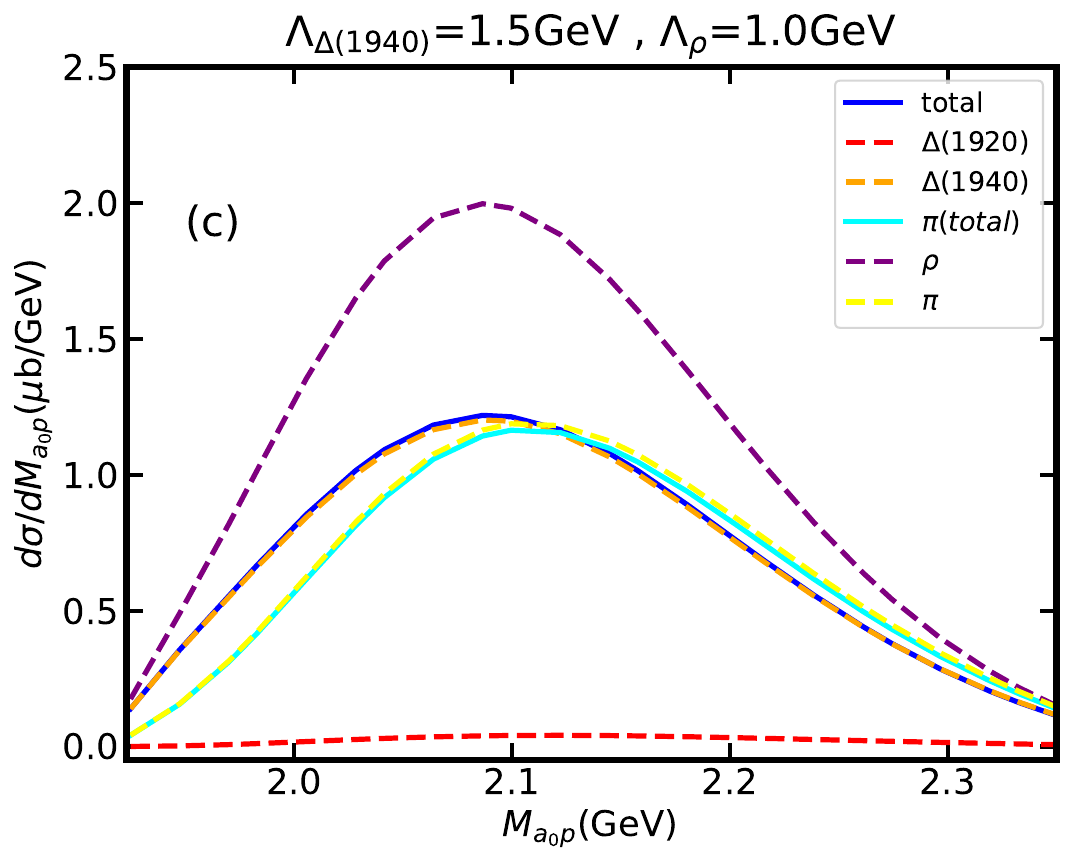}\hypertarget{c}{}
		\includegraphics[width=0.4\textwidth]{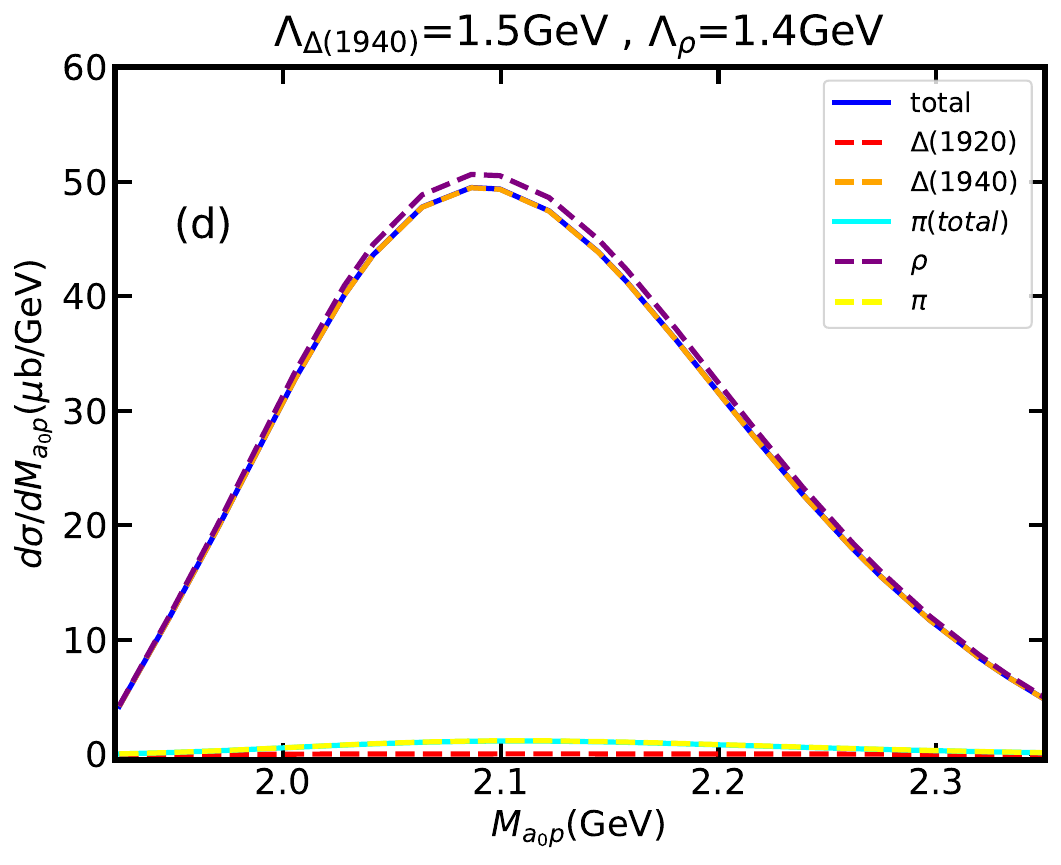}\\
		\includegraphics[width=0.4\textwidth]{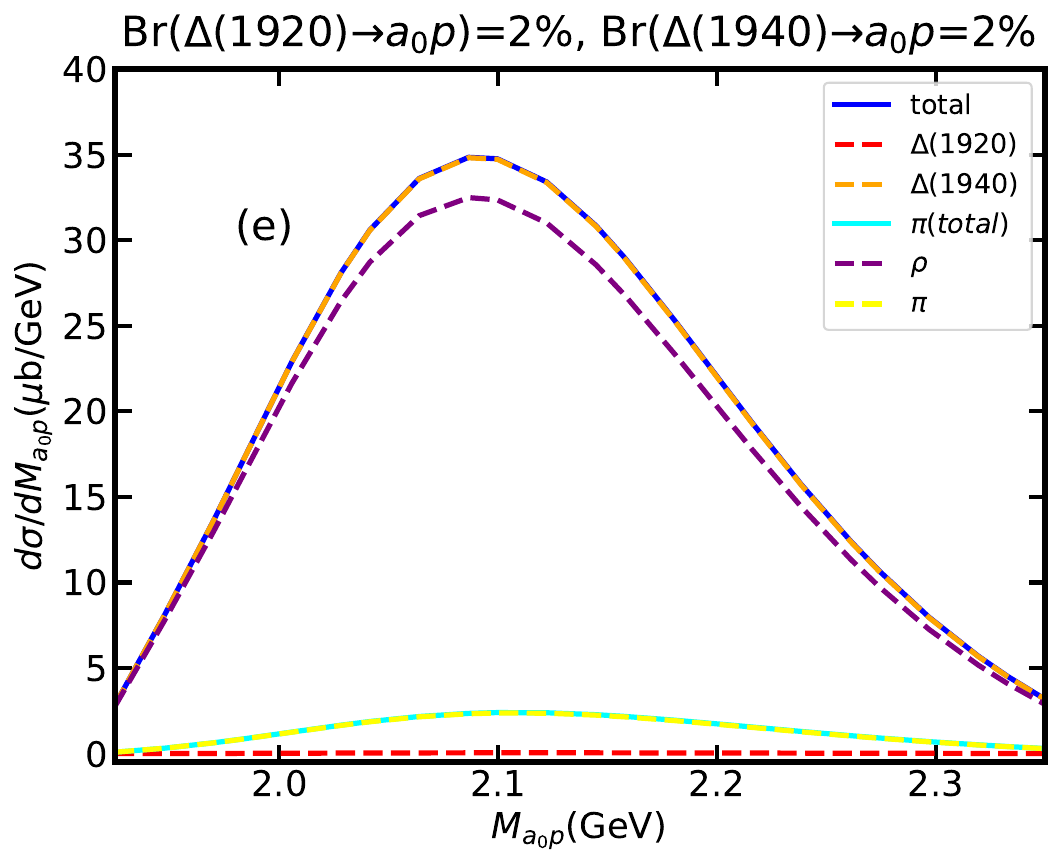}\hypertarget{e}{}
		\includegraphics[width=0.4\textwidth]{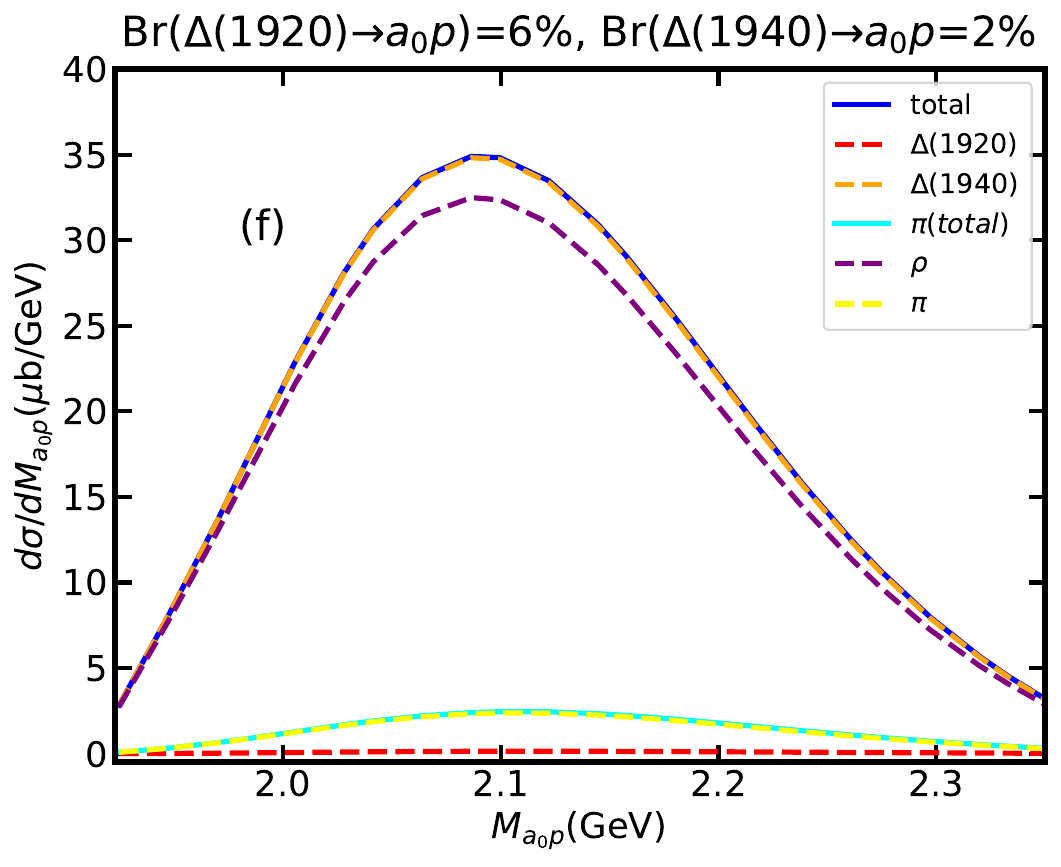}\\
		\includegraphics[width=0.4\textwidth]{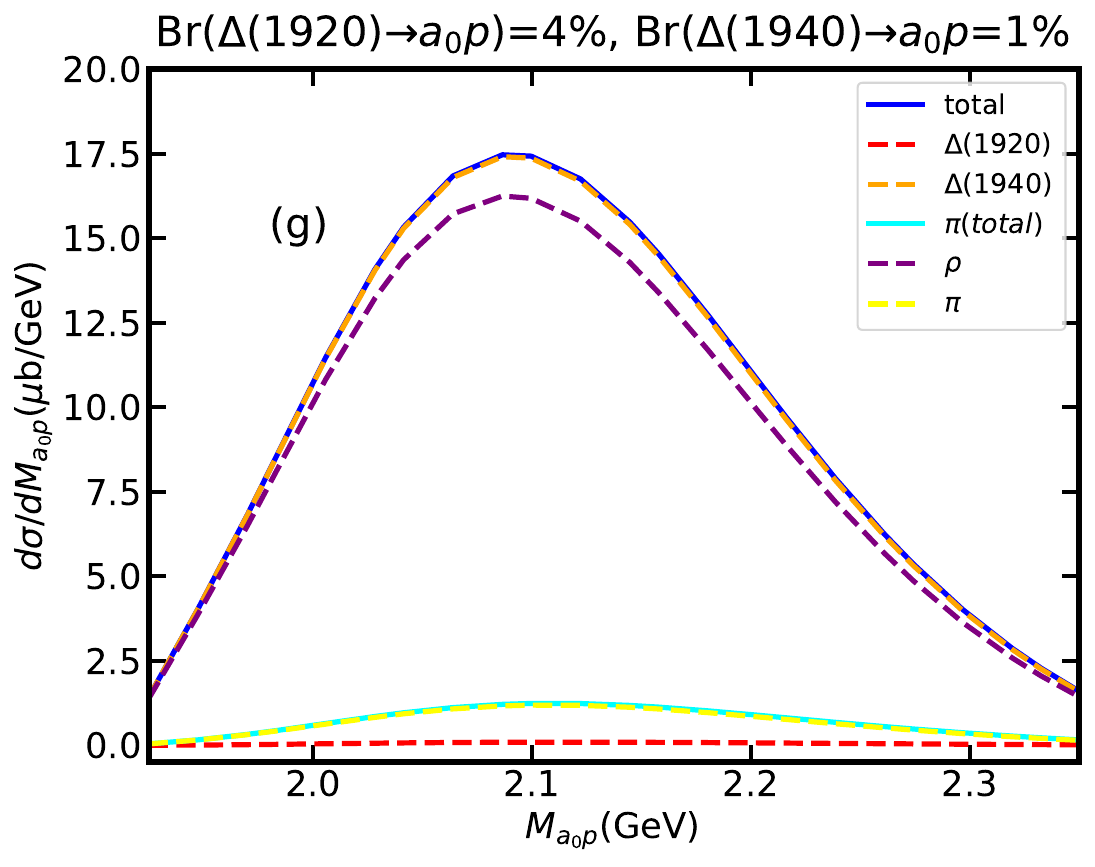}
		\includegraphics[width=0.4\textwidth]{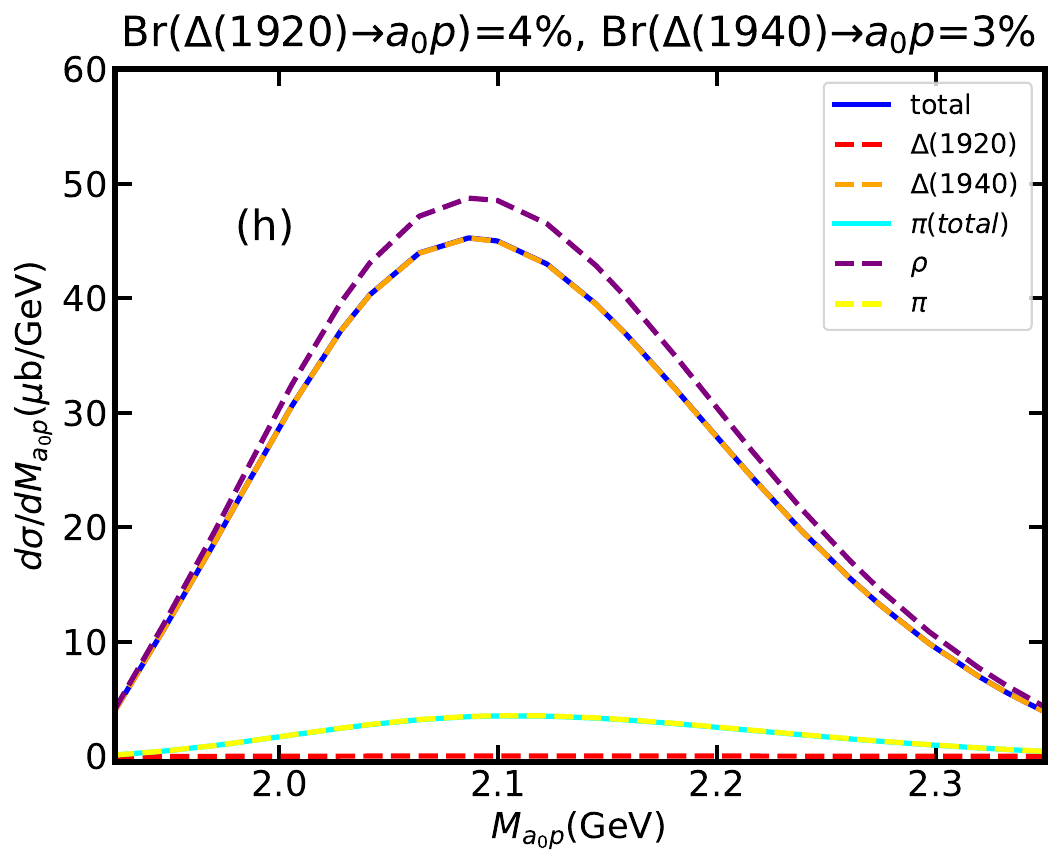}
		\captionsetup{justification=raggedright}
		\caption{At center of mass energy $W=3.5$~GeV, the invariant mass distribution of final $a_0 p$ pair for $\Sigma^{+} p\to\Lambda a_0^{+} p$ reaction under different conditions, where from (a)-(b) focus on the effect of $\Lambda_{\Delta(1940)}$; from (c)-(d) to the fourth figure focus on the effect of $\Lambda_\rho$. From (e)-(h) to the eighth figure show the changes in cross-section contributions when varying the decay branching ratio Br$(\Delta^{\ast} \to a_0 p)$.}
		\label{9}
	\end{figure*}
	Similarly, we also need to consider the dependence of the peak position in the $a_0p$ invariant mass distribution on the cutoff parameters $\Lambda_{\Delta(1940)}$ and $\Lambda_{\rho}$. In Figs.~\hyperlink{9a}{9(a)}-~\hyperlink{9d}{9(d)}, we present the results of the $a_0p$ invariant mass distribution, with $\Lambda_{\Delta(1940)}$ varying between 1.2 and 1.8~GeV, and $\Lambda_{\rho}$ varying between 1.0 and 1.4~GeV. The Fig.~\ref{9} indicate that the contribution from $\Delta(1920)$ is not significant, and $\Delta(1940)$ is the main contributor. The peak position of the total contribution is relatively insensitive to the cutoff parameters, but the curve shape of the total contribution does change with variations in the cutoff parameters. Among them, the contribution from $\rho$-meson exchange is the main contribution to $\Delta(1940)$. When the value of $\Lambda_{\rho}$ decreases, the contribution from $\rho$-exchange significantly diminishes, thereby reducing the contribution of $\Delta(1940)$. Varying only the value of $\Lambda_{\Delta(1940)}$ has no significant effect on the peak position, merely slightly altering the peak height. Changing only the value of $\Lambda_{\rho}$ results in significant changes in the peak height and the shape of the curves, but it does not alter the status of $\Delta(1940)$ and $\rho$ as the main contributors.
	
	Next, we consider the influence of the peak position in the $a_0p$ invariant mass distribution on the branching ratio of the $\Delta^{\ast} \to a_0p$ decay, as shown in Figs.~\hyperlink{9e}{9(e)}-~\hyperlink{9h}{9(h)}. Due to the significant contribution from $\rho$-meson exchange, the total contribution is still completely dominated by the $\Delta(1940)$ resonance. Regardless of whether the branching ratio for $\Delta(1920) \to a_0p$ takes its maximum or minimum value, there is almost no significant change in the peak position and curve shape. As the branching ratio for $\Delta(1940) \to a_0p$ increases, there is a notable change in the peak height, and the contribution from $\rho$-meson exchange significantly increases. In the $a_0p$ invariant mass distribution, although the peak position of the total contribution depends on the choice of cutoff parameters and decay branching ratios, the overall impact is not significant, and it does not affect the dominant position of $\Delta(1940)$.   
	
	\section{conclusion}
	We conducted a theoretical study on the process $\Sigma^{+} p \to\Lambda a_0^{+} p$ based on an effective Lagrangian approach. This model encompasses the excitation of intermediate states leading to $\Delta(1920)$ via $\pi^{+}$ and $K^{+}$ meson exchanges between the initial $\Sigma^{+}$ baryon and the initial proton $p$, as well as the production of $\Delta(1940)$ through $\pi^{+}$ and $\rho^{+}$ meson exchanges. We present predictions for the total and differential cross sections, and discuss the potential impacts of cutoff parameters, off-shell effects, and branching ratios on the $\Delta^{\ast} \to a_0 p$ decay. According to our results, the $\Delta(1940)$ resonance makes a significant contribution near the threshold, rendering this reaction suitable for investigating the characteristics of the $\Delta(1940)$ resonance and the coupling at the $\Delta(1940)Na_0$ vertex. In the $m_{a_0p}$ invariant mass distribution, the cutoff parameters and branching ratios have little influence on the peak position but can affect the peak height. Regardless of the variations in cutoff parameters and branching ratios, the peak position of the total contribution remains very close to that of the sole $\Delta(1940)$ contribution. Off-shell effects do not hinder the emergence of resonance peaks. However, cutoff parameters such as $\Lambda_{\Delta^{\ast}}$ and $\Lambda_{\rho}$, as well as the off-shell parameter Z, are generally regarded as free parameters and thus require more experimental data for determination. Given the dominance of $\Delta(1940)$ in the region close to the reaction threshold, this reaction is considered an ideal platform for delving into the unique properties of the $\Delta(1940)$ resonance. Through this approach, we can gain a more precise understanding of the intrinsic nature of this particle.
	
	\section{acknowledgments}
	This work is partly supported by the National Natural Science Foundation of China under Grants No. 12205002,
	partly supported by the the Natural Science Foundation of Anhui Province (2108085MA20,2208085MA10), 
	and partly supported by the key Research Foundation of Education Ministry of Anhui Province of China (KJ2021A0061).
	
	\bibliographystyle{unsrt}
	\bibliography{c}
\end{document}